\documentclass[11pt,a4paper]{article}
\usepackage{amsmath}
\usepackage{amssymb}
\usepackage{bm}
\usepackage{color}

\usepackage{hyperref}
\usepackage{graphicx}

\catcode `@=11 \@addtoreset{equation}{section} \catcode `@=12

\def\Tr{{\rm Tr}}

\begin{document}
\setlength{\oddsidemargin}{0cm}
\setlength{\baselineskip}{7mm}

\begin{titlepage}

	\begin{center}
		{\LARGE
		Universal Bounds on Quantum Mechanics \\
		through Energy Conservation and the Bootstrap Method
		}
	\end{center}
	\vspace{0.2cm}
	\baselineskip 18pt 
	\renewcommand{\thefootnote}{\fnsymbol{footnote}}

	\begin{center}

		Takeshi {\sc Morita}$^{a,b}$\footnote{%
			E-mail address: morita.takeshi(at)shizuoka.ac.jp
		}

		\renewcommand{\thefootnote}{\arabic{footnote}}
		\setcounter{footnote}{0}
		
		\vspace{0.4cm}
		
		{\it
			a. Department of Physics,
			Shizuoka University \\
			836 Ohya, Suruga-ku, Shizuoka 422-8529, Japan 
			\vspace{0.2cm}
			\\
			b. Graduate School of Science and Technology, Shizuoka University\\
			836 Ohya, Suruga-ku, Shizuoka 422-8529, Japan
			\vspace{0.2cm}
		}

	\end{center}
	
	
	\vspace{1.5cm}
	
	\begin{abstract}
		The range of motion of a particle with certain energy $E$ confined in a potential is determined from the energy conservation law in classical mechanics.
The counterpart of this question in quantum mechanics can be regarded as what the possible range of the expectation values of the position operator $ \langle x \rangle$ of a particle, which satisfies $E= \langle H \rangle$.
This range depends on the state of the particle, but the universal upper and lower bounds, which is independent of the state, must exist.
In this study, we show that these bounds can be derived by using the bootstrap method.
We also point out that the bootstrap method can be regarded as a generalization of the uncertainty relations, and it means that the bounds are determined by the uncertainty relations in a broad sense.
Furthermore, the bounds on possible expectation values of various quantities other than position can be determined in the same way.
However, in the case of multiple identical particles (bosons and fermions), we find some difficulty in the bootstrap method.
Because of this issue, the predictive power of the bootstrap method in multi-particle systems is limited in the derivation of observables including energy eigenstates.
In addition, we argue an application of the bootstrap method to thermal equilibrium states.
We find serious issues that temperature and entropy cannot be handled.
Although we have these issues, we can derive some quantities in micro-canonical ensembles of integrable systems governed by generalized Gibbs ensembles.

	\end{abstract}
	
\end{titlepage}

\tableofcontents

\section{Introduction}
\label{sec-intro}

In classical mechanics, it is a simple problem to find the range of motion of a particle confined in a potential.
For simplicity, we consider an one-dimensional non-relativistic particle with the mass 1 in a potential $V(x)$.
If the particle has energy $E$,
the turning points $x_1$ and $x_2$ $(x_1 < x_2)$ that satisfy $E=V(x_i)$ ($i=1,2$)  would be determined through the energy conservation law
\begin{align} 
E=\frac{1}{2}p^2+V(x),
\end{align}
and the range is given by $x_1 \le x \le x_2$.

What is the answer to this question in quantum mechanics?
One answer is that, since the particle can pass through the potential in quantum mechanics, the possible range of the position $x$ of the particle is $-\infty \le x \le \infty$.
However, this answer is not practical because the probability of taking such a large range would be exponentially small.
So, the counterpart of this problem in quantum mechanics would be
 ``If a particle satisfies $E= \langle H \rangle$, what is the possible range of the expectation value of the particle's position operator $x$?'' 
This range would depends not only on the energy but also on the state of the particle.
However, there must be some universal upper and lower bounds that are independent of the state.
In this paper, we study these bounds.
Similar questions can be asked for the expectation values of various observables.
If the system has a certain energy, how are the maximum and minimum bounds of these expectation values determined?

The flavor of this problem may be similar to that of the uncertainty relations \cite{Heisenberg1927, PhysRev.34.163}.
The uncertainty relations state that there are some universal restrictions between observables (especially variances).
In the case of the above problem, we seek the universal restrictions of observables under the additional constraint that the system has energy $E$.
Thus, the uncertainty relation might play some role. 

Actually, we can easily show that the problem in harmonic oscillators can be solved by using the uncertainty relation.
However, this method cannot be applied to more general potential cases.
There, we may need some generalization of the uncertainty relations \cite{Heisenberg1927, PhysRev.34.163} involving higher moment operator $x^n$. (See Ref.~\cite{Simon:1997es, doi:10.1080/09500340500418815, Ivan:2012zz, Li:2020vqg} for related generalizations of the uncertainty relations.)
In this paper, we point out that the bootstrap method studied by Han et al \cite{Han:2020bkb}, which was originally proposed as a new method to derive the spectrum of the energy eigenstates in quantum mechanics, can be regarded as such a generalization of the uncertainty relations.
Then, by using the bootstrap method, we numerically find the bounds in general potential cases.
Therefore, the problem is indeed closely related to the uncertainty relations.
(It means that the original bootstrap method \cite{Han:2020bkb} may also be interpreted as a derivation of the spectrum of the energy eigenstates by applying the generalized uncertainty relations.)

This problem can be asked to multi-particle systems too.
However, we find that the bootstrap method has difficulty in handling the statistical nature of identical particles (bosons and fermions) and it provides only a limited answer.
Related to this issue, the predictive power of the numerical bootstrap method for the energy eigenstates is also limited, if identical particles are involved.

Then a natural question is whether our method works in quantum many-body systems.
Particularly, the constraint $E= \langle H \rangle$ is similar to the condition for the micro-canonical ensemble in statistical mechanics, and it is valuable to apply the bootstrap method to thermal equilibrium states.
In fact, it was shown in \cite{Aikawa:2021eai} that the numerical bootstrap method works in a quantum mechanics with a sign problem, and hence the method has a potential to play a complementary role to the Monte-Carlo method in quantum many-body systems in thermal equilibrium.

However, we find a serious issue that the bootstrap method cannot handle temperature and entropy.
In addition, due to the issue of the identical particles, the convergence of the bootstrap method will not be good.
Therefore, the bootstrap method for thermal equilibrium systems may not be as good as the Monte-Carlo method.
On the other hand, as an exception, we show that,  when the system is integrable, the convergence in micro-canonical ensembles is good.
Therefore, the bootstrap method may not be useless for thermal equilibrium states in quantum many-body systems.
\\

The organization of this paper is as follows.
In section \ref{sec-one-dim}, we study the problem of finding the bounds on the expectation values of observables under the constraint $E= \langle H \rangle$ in one-dimensional quantum mechanics.
We show that this problem can be solved by using the uncertainty relation in harmonic oscillators.
For general potential cases, we can use the numerical bootstrap method to solve the problem.
We also show that the bootstrap method can be regarded as a generalization of the uncertainty relations.
In section \ref{sec-2-bdy}, two-particle systems is considered.
There, we argue that the bootstrap method has an issue on identical particles (bosons and fermions), and its predictive power is limited.
In section \ref{sec-thermal}, we show our attempt to apply the bootstrap method to thermal equilibrium states in multi-particle systems.
We see that the bootstrap method has a serious issue that  temperature and entropy cannot be handled. 
We also show that the bootstrap method in integrable systems can predict quantities except temperature and entropy.
Section \ref{sec-discussion} contains conclusions and discussions.

\section{Bootstraping One-Dimensional Particle}
\label{sec-one-dim}

\subsection{Bounds on expectation values in quantum mechanics}
\label{sec-general}

We study how the upper and lower bounds on the expectation value $\langle Q \rangle$ of an operator $Q$ are determined in quantum mechanics when the system satisfies $E= \langle H \rangle$.
We start from an one-dimensional quantum mechanics,
\begin{align} 
	H=\frac{1}{2}p^2+V(x).
	\label{H-general}
\end{align}
Here we assume $V(x) \to +\infty$ ($x\to \pm \infty$).
Our goal is to find the maximum (minimum) value of $\langle Q \rangle $ among all possible mixed states that satisfy the constraint $E=\langle H \rangle $.
Even if all the energy eigenstates of this system are known, this is a non-trivial question.\footnote{Even if we restrict the state to pure states, this problem is still non-trivial. 
In this case, a pure state is given by $\sum b_n | n \rangle$, where $  | n \rangle$ is the energy eigenstate and $b_n$ is a complex number.
Then, our task is finding a set of parameter $\{b_n \} $ such that $\langle Q \rangle $ is maximized (minimized) under the constraint $\langle H \rangle =E$.
This is a non-linear optimization problem with respect to  $\{b_n \} $, which is difficult to solve in general.
}

\begin{figure}
	\begin{center}
		\includegraphics[scale=0.47]{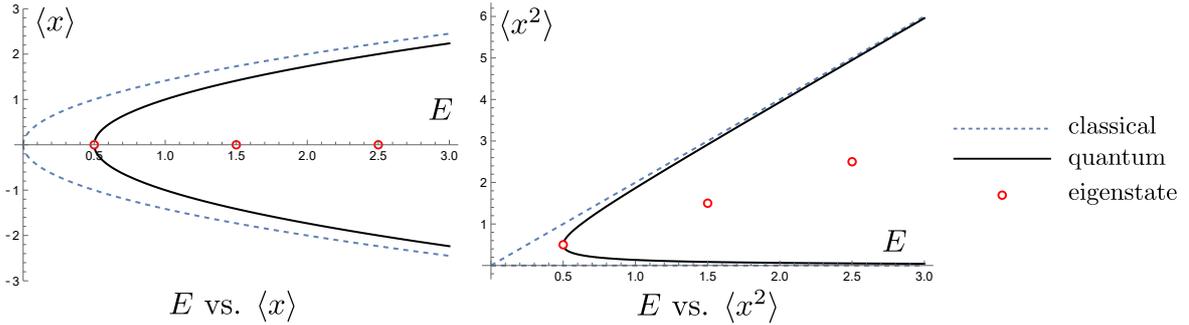}
	\end{center}
	\caption{
		The bounds on $ \langle x \rangle $ and $ \langle x^2 \rangle $ in the harmonic oscillator \eqref{H-harmonic} through the uncertainty relation. 	
		The regions enclosed by the dashed lines are the predictions from classical mechanics, and those enclosed by the solid lines are those from quantum mechanics derived in \eqref{x-bound-HO} and \eqref{x2-bound-HO}.
		The red circles are the energy eigenstates. 
		The regions in quantum mechanics are narrower than classical mechanics because of the restriction through the uncertainty relation. 
		In these plots, we have taken $\hbar=1$.
	}
	\label{Fig-HO}
\end{figure}

The flavor of this problem is similar to that of the uncertainty relations, and it is natural to employ them to find the bounds on $\langle Q \rangle$.
In fact, this attempt works for harmonic oscillators.
Let us consider the following model,
\begin{align}
	H=\frac{1}{2}p^2+\frac{1}{2} x^2.
	\label{H-harmonic}
\end{align}
First, we take $Q=x$ and investigate its bounds.
Through the constraint $E =\langle H \rangle$, we obtain
\begin{align}
E&=\frac{1}{2}\langle p^2 \rangle +\frac{1}{2} \langle  x^2 \rangle
=\frac{1}{2}\left( \langle \Delta p^2 \rangle +\langle p \rangle^2 \right)+\frac{1}{2}\left( \langle \Delta x^2 \rangle +\langle x \rangle^2 \right) \nonumber \\
& \Longrightarrow  \langle x \rangle^2+\langle p \rangle^2 = 2E - \left( \langle \Delta x^2 \rangle + \langle \Delta p^2 \rangle  \right) \ge 2E -2 \sqrt{ \langle \Delta x^2 \rangle  \langle \Delta p^2 \rangle  } 
\ge 2E - \hbar.
\label{constraint-HO-x}
\end{align}
Here, $\langle  \Delta O^2  \rangle:=\langle O^2 \rangle - \langle O \rangle^2$ denotes the deviation of $O$.
We used the arithmetic and geometric means in the first inequality, and used the uncertainty relation $ \langle \Delta x^2 \rangle \langle \Delta p^2 \rangle \ge  \hbar^2/4$ in the second inequality.
From this equation, the bounds are derived,
\begin{align} 
-x_* (E)\le  \langle x\rangle \le x_*(E), \qquad x_*(E):=\sqrt{2(E - \hbar/2)}.
\label{x-bound-HO}
\end{align}
We compare this result with the classical mechanics. In the classical mechanics, the possible range of $x$ is given by $|x| \le \sqrt{2E}$. 
Thus, if we replace $E \to E - \hbar/2$ in this relation, the quantum bounds \eqref{x-bound-HO} are reproduced.
Since $\hbar/2$ is the zero-point energy of the harmonic oscillator, this result implies that the range of $\langle x\rangle$ in quantum mechanics is narrowed by the zero-point energy. 
As $E$ increases, the difference between quantum mechanics and classical mechanics becomes relatively small, and it explains why $|x| \le \sqrt{2E}$ works in the classical limit.
These are illustrated in Fig.~\ref{Fig-HO}.

Similarly, we can derive the bounds on $Q=x^2$ through the uncertainty relation, 
\begin{align}
	2E&=2\langle H \rangle =\langle p^2 \rangle +\langle  x^2 \rangle
	\ge  \frac{\hbar^2}{4 \langle  x^2 \rangle} + \langle  x^2 \rangle  \nonumber \\
& \Longrightarrow 
E - \sqrt{E^2-\hbar^2/4} \le \langle  x^2 \rangle \le E + \sqrt{E^2-\hbar^2/4}.
\label{x2-bound-HO}
\end{align}
In classical mechanics, the range of the possible value of $\langle  x^2 \rangle =\langle  x \rangle^2 =x^2 $ is given by $ 0 \le  x^2  \le 2E$.
Thus, the result in quantum mechanics \eqref{x2-bound-HO} is again narrower than the classical one. 
It can be regarded as a consequence of the uncertainty relation, which restricts the range of physical quantities more than classical mechanics.

So far, we have investigated the upper and lower bounds on $\langle x \rangle $ and $\langle x^2 \rangle $.
Similarly, we can also obtain the bounds on $\langle p \rangle $ and $\langle p^2 \rangle $.
However, we have not shown whether states saturating these inequalities really exist or not. If not, stronger bounds must exit.
In fact, from a simple consideration shown in Appendix \ref{app-HO-x} and \ref{app-HO-x2}, we can show that coherent states saturate the bounds on $\langle x \rangle $ and $\langle p \rangle $ and that certain Gaussian wave packets saturate the bounds on $\langle x^2 \rangle $ and $\langle p^2 \rangle $. 
Hence, \eqref{x-bound-HO} and \eqref{x2-bound-HO} are the genuine bounds.
In addition, from similar considerations, we can derive the upper and lower bounds on  $\langle p \rangle $ in general potentials $V(x)$ in \eqref{H-general}.
The details are discussed in Appendix \ref{app-p}.

In this subsection, we have studied the bounds on the operators $x$, $x^2$, $p$ and $p^2$ in the harmonic oscillator \eqref{H-harmonic}. 
These bounds are restricted by the uncertainty relation.
This is in contrast to classical mechanics, where the range of physical quantities is determined only through the energy conservation law.
However, if we consider more complicated operators such as $Q=x^4$ or general potential $V(x)$ in \eqref{H-general}, it seems to be difficult to obtain the bounds on $\langle Q \rangle$ from simple uncertainty relations.

\subsection{Bootstrap analysis}
\label{sec-bootstrap}

To find the bounds in more general situations, we apply the numerical bootstrap method proposed by Han et al \cite{Han:2020bkb}.
As we discuss soon, this method can be regarded as a generalization of the uncertainty relations.

For this purpose, we follow Han et al and introduce a bootstrap matrix.
We assume that any (non-singular) operators $O$ in the system satisfy the following positivity condition for any (non-singular) mixed states 
\begin{align}
	\langle O^\dagger O \rangle_\rho := \Tr \left( \hat{\rho} O^\dagger O \right) \ge 0.
	\label{positivity}
\end{align}
Here $\hat\rho$ is defined by
\begin{align}
	\hat \rho := \sum_\alpha c_\alpha | \alpha \rangle \langle \alpha |,
\end{align}
where $| \alpha \rangle $ are pure states that are normalized as $ \langle \alpha | \alpha \rangle =1 $, and $c_\alpha$ are real constants that specify the mixed state and satisfy 
$0 \le c_\alpha  \le 1$ and $\sum_\alpha c_\alpha =1$.
We also assume that $O$ satisfies,
\begin{align} 
 \langle O^\dagger \rangle_\rho = \langle O \rangle^*_\rho.
 \label{conjugate}
\end{align}

Then, we prepare a set of some $K$ operators $\{ O_n \}$ and $K$ auxiliary constants $\{ b_n \}$  $(n=1,\cdots,K )$, and define an operator $\tilde{O}$,
\begin{align}
	\tilde{O}:= \sum_{n=1}^K b_n O_n.
	\label{Otilde}
\end{align}
Now, because of the positivity condition \eqref{positivity}, 
\begin{align}
	\langle   \tilde{O}^\dagger \tilde{O}  \rangle_\rho 
	=
	\sum_{m,n=1}^K  b^{*}_m b_n	\langle   O_m^\dagger O_n  \rangle_\rho 
	\ge 0
	\label{positive-Otilde}
\end{align}
is satisfied for arbitrary constants $\{ b_{n} \}$.
Hence, the following $K \times K$ Hermite matrix ${\mathcal M}$ has to be positive-semidefinite \cite{Han:2020bkb},
\begin{align}
	{\mathcal M}:=
	\begin{pmatrix}
		\left\langle  O_1^{\dagger} O_1 \right\rangle_\rho & \left\langle  O_1^{\dagger} O_2  \right\rangle_\rho & \cdots &\left\langle  O_1^{\dagger} O_K  \right\rangle_\rho \\
		\left\langle  O_2^{\dagger} O_1  \right\rangle_\rho & \left\langle O_2^{\dagger} O_2  \right\rangle_\rho & \cdots &\left\langle  O_2^{\dagger} O_K  \right\rangle_\rho \\
		\vdots                                                         & \vdots                                                         & \ddots & \vdots \\
		\left\langle  O_K^{\dagger} O_1  \right\rangle_\rho  & \left\langle  O_K^{\dagger} O_2  \right\rangle_\rho & \cdots &\left\langle  O_K^{\dagger} O_K \right\rangle_\rho
	\end{pmatrix}
	\succeq 0,
	\label{bootstrap}
\end{align}
where $\succeq $ is the mathematical symbol for a positive-semidefinite matrix.
We call ${\mathcal M}$ as a bootstrap matrix and $\tilde{O}$ as its seed operator.
We later see that $K$ may be regarded as a cut off parameter of numerical bootstrap analysis.

The discussion up to this point is not limited to one-dimensional quantum mechanics and it can be applied to general systems.
From now on, we focus on one-dimensional quantum mechanics with the Hamiltonian  \eqref{H-general}.
Here we take the seed operator 
\begin{align}
	\tilde{O}= \sum_{m=0}^{K_x} \sum_{n=0}^{K_p} b_{mn} x^m p^n,
	\label{Otilde-xp}
\end{align}
and construct the bootstrap matrix from it,
\begin{align}
	{\mathcal M}=
	\begin{pmatrix}
		1 & \left\langle x \right\rangle_\rho  & \left\langle p \right\rangle_\rho   & \cdots  \\
		\left\langle x	\right\rangle_\rho & \left\langle x^2	\right\rangle_\rho & \left\langle xp \right\rangle_\rho &  \cdots  \\
		\left\langle p	\right\rangle_\rho  &  \left\langle px \right\rangle_\rho & \left\langle p^2 \right\rangle_\rho  & \cdots   \\
		\vdots & \vdots  & \vdots  & \ddots \\
	\end{pmatrix}.
	\label{bootstrap-XP}
\end{align}
The components of this matrix take the forms $\langle p^h x^k p^l  \rangle_\rho$, and
they can be described by the ordered forms $\langle x^m p^n  \rangle_\rho$ through the relation that can be derived from the commutator relation $[x,p]=i \hbar$,
\begin{align}
	p^n x^{m} =
	\sum_{k=0}^{\min (m,n)}  (-i \hbar)^{k} \frac{n!m!}{k!(n-k)!(m-k)!}  x^{m-k} p^{n-k}.
	\label{ordering}
\end{align}
In addition, these components $\langle x^m p^n  \rangle_\rho$ are restricted from the condition \eqref{conjugate}.
For example, $\langle px \rangle_\rho=\langle xp \rangle_\rho^*=\langle xp \rangle_\rho - i \hbar$, and it implies that ${\rm Im}( \langle xp \rangle_\rho) = \hbar/2$.

\subsubsection{Bootstrap analysis and the uncertainty relations}
\label{sec-uncertainty}

The condition $	{\mathcal M} \succeq 0$ strongly constrains the possible values of the quantities $\langle x^m p^n  \rangle_\rho$.
Actually, the uncertainty relation  $\langle \Delta x^2 \rangle \langle \Delta p^2 \rangle \ge \hbar^2/4$ is one of the consequences of this condition. 
Thus, the condition ${\mathcal M} \succeq 0$ may be regarded as a generalized version of the uncertainty relations.

To see the derivation of the uncertainty relation from ${\mathcal M} \succeq 0$, we take $K_x=K_p=1$ in \eqref{Otilde} \footnote{See \cite{Simon:1997es, Curtright:2001jn} for a related derivation of the uncertainty relation.}
\begin{align}
	\tilde{O}= b_{00} 1 +  b_{10} x + b_{01} p.
	\label{Otilde-uc}
\end{align}
Then the bootstrap matrix becomes
\begin{align}
	{\mathcal M}=
	\begin{pmatrix}
	1&	\left\langle x \right\rangle & \left\langle  p \right\rangle \\
	\left\langle x \right\rangle &	\left\langle x^2 \right\rangle & \left\langle  xp \right\rangle \\
	\left\langle p \right\rangle &	\left\langle px \right\rangle  & \left\langle p^2 \right\rangle \\
	\end{pmatrix} .
\end{align}
Here we have omitted the symbol $\rho$.
This matrix should be positive-semidefinite, and the determinant is non-negative.
Thus, we obtain \cite{Sakurai:2011zz},
\begin{align}
	(\left\langle x^2 \right\rangle-\left\langle x \right\rangle^2 )(\left\langle p^2 \right\rangle-\left\langle p \right\rangle^2 ) &  \ge | \left\langle  xp \right\rangle - \left\langle  x \right\rangle\left\langle  p \right\rangle |^2= 
	\frac{1}{4} |\left\langle  [x,p] \right\rangle +  \left\langle \{x,p\} \right\rangle  - 2\left\langle  x \right\rangle\left\langle  p \right\rangle |^2 \nonumber \\
	&	=
	\frac{1}{4} |\left\langle  [x,p] \right\rangle|^2+\frac{1}{4} |\left\langle  \{x,p\} \right\rangle  - 2\left\langle  x \right\rangle\left\langle  p \right\rangle |^2
	\ge \frac{1}{4} |\left\langle  [x,p] \right\rangle|^2.
	\label{uc1}
\end{align}
Here, we have used that $\{x,p\}$ is hermitian and $[x,p] $ is anti-hermitian in the third equality.
Then, by using $[x,p]=i\hbar$, we obtain the uncertainty relation
\begin{align}
	\langle \Delta x^2 \rangle \langle \Delta p^2 \rangle
	\ge \frac{1}{4} \hbar^2.
	\label{uncertainty}
\end{align}
Therefore, the condition ${\mathcal M} \succeq 0$ for a general bootstrap matrix \eqref{bootstrap} may be regarded as an extended version of the uncertainty relations.\footnote{
	Various generalizations of the uncertainty relations involving higher moment operators have been proposed. For example, see Refs.~\cite{Simon:1997es, doi:10.1080/09500340500418815, Ivan:2012zz, Li:2020vqg}.
	Closely related inequalities can be obtained through the cumulant expansion \cite{kubo1962generalized} and the Jensen's inequality, too.
	One advantage of the constraint ${\mathcal M} \succeq 0$ in our analysis is that it can be solvable through a linear programming as we demonstrate.
	}
(Actually, this condition for $K_x \ge 1$ and $K_p \ge 1$ is stronger than the original uncertainty relation, since the uncertainty relation can be obtained from \eqref{positive-Otilde} by tuning $b_{mn}=0$ except $b_{00}$,  $b_{01}$ and $b_{10}$.)

\subsection{Bootstraping one-dimensional models with $E= \langle H \rangle$}
\label{sec-bound}

Han et al used the condition $ {\mathcal M} \succeq 0$ to obtain the spectrum of the energy eigenstates, which will be reviewed in Sec.~\ref{sec-eigen}. 
Here, we apply this condition to solve our problem of finding the upper and lower bounds on the expectation values of an operator $Q$ when the system with the Hamiltonian \eqref{H-general} satisfies $E= \langle H \rangle$.
In mathematics, this optimization problem is represented by the following symbols: 
\begin{align}
&	\max \{
	\langle Q \rangle_\rho	~|~
	{\mathcal M}
	\succeq 0 \wedge E = \langle H \rangle_\rho = \frac{1}{2} \langle p^2 \rangle_\rho +  \langle V( x) \rangle_\rho
	\} ,  \nonumber \\
	&	\min \{
		\langle Q \rangle_\rho	~|~
		{\mathcal M}
		\succeq 0 \wedge E = \langle H \rangle_\rho = \frac{1}{2} \langle p^2 \rangle_\rho +  \langle V( x) \rangle_\rho
		\}  .
	\label{constraint-general-E}
\end{align}
These are linear programming with respect to the quantities $\{ \langle x^m p^n \rangle_\rho \}$, and are solvable.

\subsubsection{Examples: Anharmonic oscillator and double-well potential}
\label{sec-general-example}

\begin{figure}
	\begin{center}
		\includegraphics[scale=0.47]{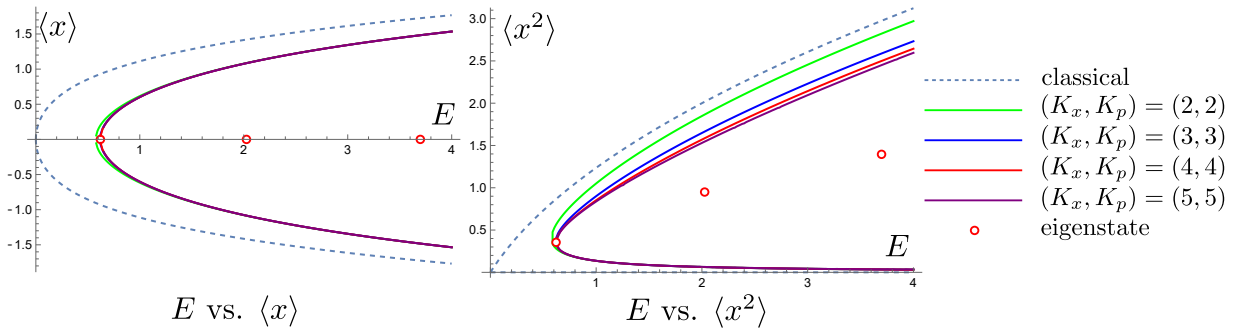}
	\end{center}
	\caption{
		The ranges of the possible expectation values of $ \langle x \rangle $ and $ \langle x^2 \rangle $ in the anharmonic oscillator \eqref{H-aho} through the bootstrap method. The expectation values can be taken in the region enclosed by the colored solid curves in the figures (the colors represent different $(K_x,K_p)$). The dashed lines are for classical mechanics, and the red circles are the quantities of the energy eigenstates. 
		As $(K_x,K_p)$ increase, the range becomes smaller, and provides the stronger bound.
		As in the case of the harmonic oscillator, the ranges in quantum mechanics are narrower than those in classical mechanics. It can also be seen that the smallest possible value of $E$ coincides with the ground state.
	}
	\label{Fig-general-aho}
\end{figure}

\begin{figure}
	\begin{center}
		\includegraphics[scale=0.47]{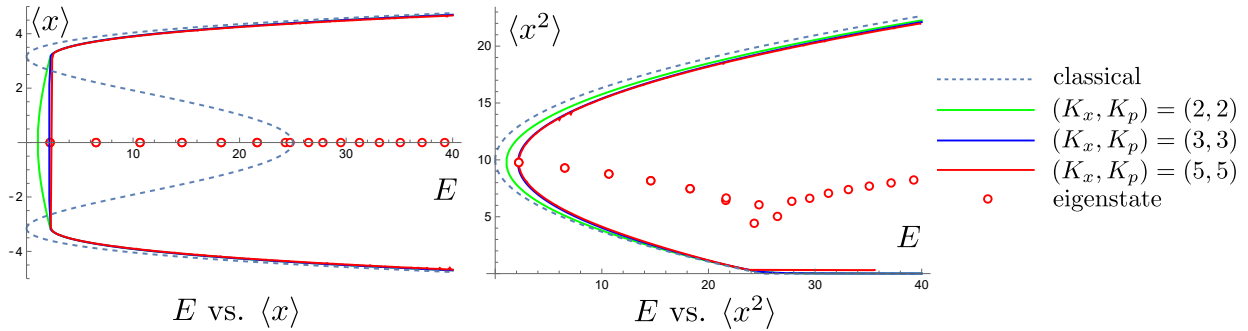}
	\end{center}
	\caption{
		The range of the possible expectation values of $ \langle x \rangle $ and $ \langle x^2 \rangle $ in the double-well potential \eqref{H-DW} through the bootstrap method. The expectation values can be taken in the region enclosed by the colored solid curves in the figures. The dashed lines are for classical mechanics, and the red circles are for the quantities of the energy eigenstates. 
		Similar to the anharmonic oscillator case shown in Fig.~\ref{Fig-general-aho}, the result converges as  $(K_x,K_p)$ increase.
		It can be seen that, in quantum mechanics, the range of the possible expectation values of $ \langle x \rangle $ also appears in the region not allowed in classical mechanics.
	}
	\label{Fig-general-DW}
\end{figure}

As examples, we investigate this problem in an anharmonic oscillator and a double-well potential,
\begin{align}
	H  =& \frac{1}{2}  p^2  + \frac{1}{2} x^2 
	   + \frac{1}{4}  x^4 ,
	\label{H-aho} \\
	 H  =& \frac{1}{2} p^2  - 5  x^2 
	+ \frac{1}{4} x^4.
	\label{H-DW} 
\end{align}
We take $Q=x$ and $Q=x^2$, and compute the maximum and minimum values of the expectation values of these operators by solving the linear programming \eqref{constraint-general-E}.\footnote{For these analyses, we used the Mathematica package ``SemidefiniteOptimization".
We use the version 13.0 and 13.1.
Note that the numerical results highly depend on the option ``Method" of this package.
The results presented in Figs.~\ref{Fig-general-aho} and \ref{Fig-general-DW} are obtained using Method $\to$ ``DSDP".
In the numerical analysis throughout this paper, $\hbar=1$ is taken.
}
The results are shown in Figs.~\ref{Fig-general-aho} and \ref{Fig-general-DW}.
We find that the numerical results converge as we take larger values of $K_x$ and $K_p$ defined in \eqref{Otilde-xp}.
The convergence is very quick for $ \langle x \rangle $, and is good enough even at $K_x=K_p=2$.
It also seems to converge reasonably quickly for $ \langle x^2 \rangle $.
These results indicate that the numerical bootstrap method is quite effective in our problem.
Here, one remark is that, similar to the uncertainty relation in the harmonic oscillator, the bootstrap method cannot tell us whether the states saturating the bounds exist or not.
Since such states are found in the harmonic oscillator, it is likely that such states exist in the current models too.

In the case of the anharmonic oscillator, the obtained results are qualitatively similar to the harmonic oscillator case shown in Fig.~\ref{Fig-HO}.
In the case of the double-well potential, $ \langle x \rangle $ can take a value in the forbidden region in the classical mechanics.\footnote{
	Even in classical mechanics, if we allow probabilistic states of a particle, $\langle x \rangle$ can take a value in the forbidden region \cite{Nakayama:2022ahr}. 
	We do not consider such states here. 
	A related study of the bootstrap method for classical particles was done in Refs.~\cite{Nakayama:2022ahr, Berenstein:2022ygg}.	}
This is expected in quantum mechanics, and our result reproduces this property.

In these numerical results, the lowest energy points coincide with the ground states.\footnote{We compute the eigenstates by numerically solving the Schr$\ddot{\rm o}$dinger equations.
We use the Mathematica package ``NDEigensystem" throughout this paper.
}
This is because the ground state is realized as the lowest energy state among all possible states.
If one simply want to find the energy of the ground state, one can  obtain it by numerically solving the linear programming: $\min\{ \langle H \rangle~|~  {\mathcal M} \succeq 0 \} $ \cite{Lawrence:2021msm}.
It implies that the ground state is the optimized state that minimizes energy under the generalized uncertainty relation ${\mathcal M} \succeq 0 $, and it might give us a new picture of the ground state in quantum mechanics.
\\

We have seen that the numerical bootstrap method works effectively in our problem.
In principle, we can apply our analysis to arbitrary operators $Q$.
However, the problem of finding the bounds on the possible values of the product of expectation values $\langle Q_1 \rangle \langle Q_2 \rangle$ would be a nonlinear optimization problem, and it would be numerically much harder.

\subsection{Bootstrapping stationary states with $E= \langle H \rangle$}
\label{sec-stationary}

We have seen that, when the system has energy $E$, the upper and lower bounds on the expectation value $\langle Q \rangle$ can be obtained by using the bootstrap method.
There, we have imposed no restrictions other than energy $E=\langle H \rangle$ on the states.
From now on, we consider the possible range of the expectation value $\langle Q \rangle$ under an additional condition that the states are stationary.
This will be a hint when we apply the bootstrap method to thermal equilibrium states, which we discuss in Sec.~\ref{sec-thermal}.
The general stationary state in quantum mechanics is given by the following mixed state $\hat \rho_{\rm st}$,
\begin{align}
	\hat \rho_{\rm st} = \sum_{n=0}^{\infty} c_n | n \rangle \langle n | ,
	\label{mixed-static}
\end{align}
where $|n \rangle$ ($n=0,1,\cdots$) is the energy eigenstate with the eigen energy $E_n$, and 
$c_n$ is a constant satisfying $0 \le c_n \le 1$ and $\sum_n c_n=1$.
Then, this state satisfies\footnote{It was pointed out in \cite{Berenstein:2022ygg} that we need to take a special care on this condition when the system is on a half line such as a radial coordinate $r\in [0,+\infty]$.}
\begin{align}
	\langle [ H,O] \rangle_{\rho_{\rm st}}=0 
	\label{HO=0}
\end{align}
for any operators $O$.
Thus, in order to obtain the bounds on $\langle Q\rangle_{\rho_{\rm st}} $ for the stationary states with energy $E=\langle H \rangle$, we should add this condition to the constraints \eqref{constraint-general-E}, and solve the optimization problem of finding the maximum and minimum values of $\langle Q\rangle_{\rho_{\rm st}} $,
\begin{align}
	{\mathcal M}
	\succeq 0, \quad E = \langle H \rangle_{\rho_{\rm st}} = \frac{1}{2} \langle p^2 \rangle_{\rho_{\rm st}} +  \langle V(x) \rangle_{\rho_{\rm st}},  \quad  	\langle [ H,O] \rangle_{\rho_{\rm st}}=0, \quad (O \in \{x^m p^n \}) .
	\label{constraint-statonary-E}
\end{align}
This is again a linear programming program with respect to $\{ \langle x^m p^n \rangle_{\rho_{\rm st}} \}$, and we can compute it numerically.

Note that, different form the problem \eqref{constraint-general-E} discussed in Sec.\ref{sec-bootstrap}, we can solve this problem if we know all the energy eigenstates of the Hamiltonian.
For example, if all $\{c_n \}$ except $c_0$ and $c_1$ in \eqref{mixed-static} are zero, the range of the possible values of $\langle Q\rangle_{\rho_{\rm st}} $ is limited to the straight line connecting the point $(E,\langle Q \rangle)$ at  $|0 \rangle$  and  $|1 \rangle$, since $c_0+c_1=1$.
Extending it to non-zero $\{c_n \}$, we will obtain the region enclosed by the polygonal line connecting the eigenstates, and $\langle Q\rangle_{\rho_{\rm st}} $ can take a value only in this region.

\subsubsection{Example: Anharmonic oscillator}
\label{sec-stationary-example}

\begin{figure}
	\begin{center}
		\includegraphics[scale=0.6]{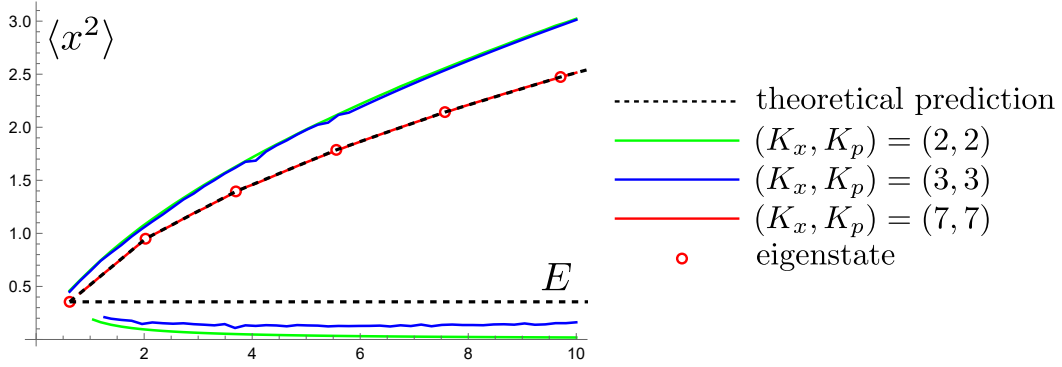}
	\end{center}
	\caption{
		The range of possible values of $\langle x^2 \rangle_{\rho_{\rm st}}$ of the anharmonic oscillator \eqref{H-aho} for stationary states \eqref{mixed-static}.
		The results of the bootstrap method \eqref{constraint-statonary-E} plotted by the colored solid lines converge to the theoretical predictions (the dashed lines) as $K_x$ and $K_p$ increase. However, the calculation of the lower bound on $\langle x^2 \rangle_{\rho_{\rm st}}$ is numerically difficult and does not give reliable answers.
		(The lower bound at $(K_x,K_p)=(7,7)$ is not shown in the figure, because we could not obtain reliable results.)
		On the other hand, the upper bound in $E <10$ converges well, and the results are almost identical with the theoretical prediction.
	}
	\label{Fig-static}
\end{figure}

As an example, we seek the bounds on $Q=x^2$ in the anharmonic oscillator \eqref{H-aho}.
(Note that $\langle x \rangle_{\rho_{\rm st}}=0$ in the anharmonic oscillator.)
We derive the energy eigenstates by numerically solving the Schr$\ddot{\rm o}$dinger equation.
The energy and  $\langle x^2\rangle $ for each eigenstate is plotted by the red circles in Fig.~\ref{Fig-static}.
Then, the range of the possible values of $\langle x^2\rangle_{\rho_{\rm st}} $ for the stationary states \eqref{mixed-static} is given by the polygonal region connecting them, which is illustrated by the dashed lines in Fig.~\ref{Fig-static}.
Since the constraints \eqref{constraint-statonary-E} are more stronger than \eqref{constraint-general-E}, 
the range of the possible values of  $\langle x^2\rangle $ is narrower than the range shown in Fig.~\ref{Fig-general-aho}.

We solve the same problem by computing \eqref{constraint-statonary-E} via the numerical bootstrap method.\footnote{
	In this numerical problem, one question is what operator $x^m p^n$ should be taken in the constraint $\langle [ H,O] \rangle_{\rho_{\rm st}}=0$ in \eqref{constraint-statonary-E}.
	If more operators of $O=x^m p^n$ are taken, the constraints become stronger, but they require greater computational resources.
	Here, we simply take the operators appearing in the bootstrap matrix ${\mathcal M}$.
	(Note that new operators $O'$, which do not exist in the bootstrap matrix ${\mathcal M}$, will appear from the equation $\langle [ H,O] \rangle_{\rho_{\rm st}}=0$, and we can derive new constraints $\langle [ H,O'] \rangle_{\rho_{\rm st}}=0$ with respect to these new operators, but we do not do it.)
	Similar questions arise in other bootstrap problems too and we take the same prescription throughout this paper.
	An exception is the energy eigenstate problem in one-dimension quantum mechanics discussed in Sec.~\ref{sec-eigen}. There, we can explicitly solve the constraints $\langle [ H,O] \rangle=0$ and $\langle HO \rangle=E\langle O \rangle$ \cite{Aikawa:2021qbl}.}
	The results are shown in Fig.~\ref{Fig-static}.\footnote{We used the Mathematica package ``SemidefiniteOptimization" and take ``CSDP" or ``MOSEK" as the option ``Method".}
We see that the predictions of the bootstrap method asymptotically approach the theoretical prediction (the dashed lines) as $K_x$ and $K_p$ increase.
In particular, the upper bound on $\langle x^2 \rangle_{\rho_{\rm st}}$ at $(K_x,K_p)=(7,7)$ is apparently almost identical to the dashed lines.
	
However, there are some regions where the numerical bootstrap method does not work well.
As energy increases, the upper bound on $\langle x^2 \rangle_{\rho_{\rm st}}$ becomes more and more difficult to obtain.
Actually, in the original bootstrap method \cite{Han:2020bkb}, the eigenstates are obtained from lower energy eigenstates \cite{Aikawa:2021qbl}.
Hence, the bootstrap method may work better in lower energy in general.
Besides, the lower bound on $\langle x^2 \rangle_{\rho_{\rm st}}$ is also difficult to obtain when the size of the bootstrap matrix size is large.
Note that the lower bound in Fig.~\ref{Fig-static} is the straight line connecting the ground state and the state at $E=\infty$.
Since handling high energy states may be difficult in the bootstrap method, it might explain why the bootstrap method does not work well for deriving the lower bound.

One interesting feature of this result is that $\langle x^2 \rangle$ at the energy eigenstates are reproduced as the vertexes of the polygonal region in Fig.~\ref{Fig-static}.
Thus, the constraints \eqref{constraint-statonary-E} is enough to obtain the energy eigenstates in the bootstrap method.
However, this is because $(E,\langle x^2 \rangle)$ for the eigenstates in the anharmonic oscillator lie on a convex curve, and it does not generally occur.
If we change the Hamiltonian and $(E,\langle x^2 \rangle)$ distribute more complicatedly, some of the eigenstates appear in the inside of the polygon and they cannot be observed.
(See, for example, $(E,\langle x^2 \rangle)$ in the double-well potential shown in Fig.~\ref{Fig-general-DW}.)
In order to find the full eigenstates through the bootstrap method, we need to add further constraints to \eqref{constraint-statonary-E} as we
 see in the next section.

\subsection{Bootstrapping energy eigenstates}
\label{sec-eigen}

We have investigated the range of the possible values of $\langle Q \rangle$ in the stationary state, and now we impose further constraints to obtain the energy eigenstates.
(This is the original bootstrap method proposed by Han et al \cite{Han:2020bkb}.)
The energy eigenstate $|E \rangle$ satisfies not only the stationary condition \eqref{HO=0} but also the following equation
\begin{align}
	 & \langle E| HO | E \rangle=E \langle E| O|E \rangle
	\label{HO=EO}
\end{align}
for any well defined operators $O$. 
Hence, we may obtain the spectrum of the energy eigenstates by evaluating the possible values of $\langle Q \rangle$ under the constraints,
\begin{align}
	{\mathcal M}
	\succeq 0, \quad 
	\langle E|   HO  |E \rangle =E\langle E| O  |E \rangle, \quad \langle E|  [ H,O] |E \rangle=0,
	\quad (O \in \{x^m p^n \}).  
	\label{constraint-eigen}
\end{align}
Here the constraint $E= \langle E| H | E \rangle $ in \eqref{constraint-statonary-E} is involved in \eqref{HO=EO} with $O=1$.
The constraints \eqref{HO=0} and \eqref{HO=EO} are quite strong, and they reduce to the following recurrence relation \cite{Aikawa:2021qbl,  Berenstein:2021dyf, Bhattacharya:2021btd, Hu:2022keu},
\begin{align}
	n(n-1)(n-2) \langle x^{n-3} \rangle -8n \langle x^{n-1} V(x)\rangle +8 n E \langle x^{n-1}  \rangle -4 \langle x^{n} V'(x)\rangle=0.
	\label{AHO-recurrence}
\end{align}
Here we have omitted $|E \rangle$.
When $V(x)$ is a polynomial, this recurrence relation can be solved and $\langle x^n \rangle$ for any integers $n$ are expressed by a finite number of operators $\{ \langle x^m \rangle \}$ and $E$.
Similarly, $\langle x^k p^l \rangle$ for any integers $k$ and $l$ is also described by these quantities \cite{Aikawa:2021qbl}.
Previous studies have shown that these conditions are strong enough to reproduce the observables in the energy eigenstates \cite{Han:2020bkb, Aikawa:2021eai, Berenstein:2022ygg, Aikawa:2021qbl, Berenstein:2021dyf, Bhattacharya:2021btd, Berenstein:2021loy, Tchoumakov:2021mnh, Du:2021hfw, Bai:2022yfv, Li:2022prn, Khan:2022uyz}.

\subsubsection{Example:  Anharmonic oscillator}
\label{sec-eigen-example}

\begin{figure}
	\begin{center}
		\includegraphics[scale=0.6]{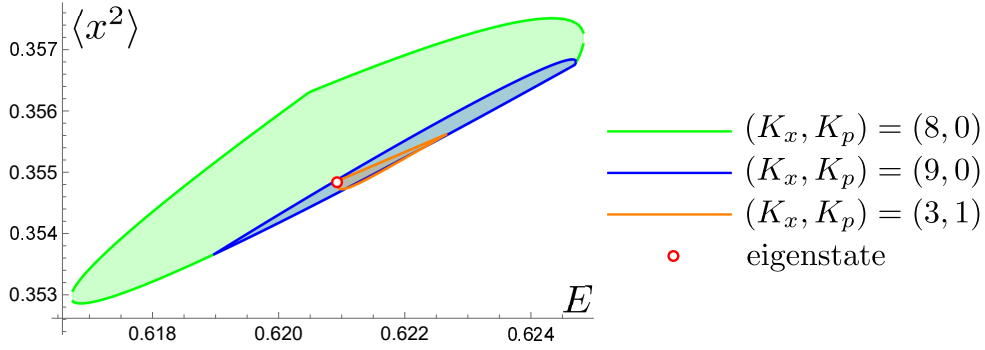}
	\end{center}
	\caption{
		The possible ranges of $\langle x^2 \rangle$ for the energy eigenstates in the anharmonic oscillator \eqref{H-aho} near the ground state.
		The numerical bootstrap problem with the constraints \eqref{constraint-eigen} is solved.
		The red circle is the result of numerically solving the Schr$\ddot{\rm o}$dinger equation, and the results of the bootstrap method asymptotically approach to it. Similar results can be obtained for other energy eigenstates too, but it works better for lower energies \cite{Aikawa:2021qbl}.
	}
	\label{Fig-eigen}
\end{figure}

We demonstrate the derivation of the energy eigenstates in the anharmonic oscillator \eqref{H-aho}.
In this case, by solving the recurrence relation \eqref{AHO-recurrence}, the operator $\langle x^k p^l \rangle$ is expressed by $\langle x \rangle$, $\langle x^2 \rangle$ and $E$.
Then, the bootstrap matrix \eqref{bootstrap-XP} becomes
\begin{align}
	{\mathcal M}=
	\begin{pmatrix}
		1 & \left\langle x	\right\rangle  & 0   & \cdots  \\
		\left\langle x	\right\rangle  &  \left\langle x^2	\right\rangle & \frac{i}{2} &  \cdots  \\
		0  &  -\frac{i}{2} & \frac{1}{3}\left( 4E- \left\langle x^2	\right\rangle \right)  & \cdots   \\
		\vdots & \vdots  & \vdots  & \ddots \\
	\end{pmatrix}.
	\label{bootstrap-XP2}
\end{align}
The constraint ${\mathcal M} \succeq 0$ for this matrix is quite strong and the allowed regions are point-like.
See Fig.~\ref{Fig-eigen}.\footnote{
	Our analysis is the same as that of Han et al, with two differences.
	One is that Han et al took $K_p=0$ in \eqref{Otilde-xp} but we did not.
	Actually, $K_p \neq 0$ might improve the numerical analysis \cite{Aikawa:2021qbl}.
	Another difference is that Han et al. imposed $\langle x \rangle=0$ by hand, but we did not, although  $\langle x \rangle=0$ reduces the computational resources.
	This is because we want to emphasize that the constraint \eqref{constraint-eigen} is sufficient to obtain the eigenstates. Note that our obtained states satisfy $\langle x \rangle \simeq 0$.}~\footnote{
	The bootstrap matrix linearly depends on $\langle x \rangle$ and $\langle x^2 \rangle$ and non-linearly depends on $E$.
	Thus, if we fix $E$, the optimization problem \eqref{constraint-eigen} can be solved by using a linear programming.
	We solve this problem by using the Mathematica package ``SemidefiniteOptimization" and take ``DSDP" as the option ``Method".}

\subsection{Summary of the one-dimensional problem}
\label{sec-summary-1d}

Let us summarize the discussions in this section.
The bootstrap method allows us to obtain the range of the possible values of observables by applying the various constraints to the expectation values.
In particular, the constraints used in this section have the following meanings:
\begin{itemize}
	\item  ${\mathcal M} \succeq 0$: A generalization of the uncertainty relations.
	\item   $E=\langle H \rangle$: The state has energy $E$.
	\item $\langle [ H,O] \rangle=0$: The state is stationary. 
	\item $\langle [ H,O] \rangle=0$ and $\langle HO \rangle=E\langle O \rangle$: The state is the energy eigenstate with energy eigenvalue $E$.
\end{itemize}
The first constraint ${\mathcal M} \succeq 0$ should be satisfied always, and the rest of the constraints specify the states, which we want to investigate.


\section{Bootstrapping Two-Particle Systems}
\label{sec-2-bdy}

\subsection{Problems in identical particles}
\label{sec-identical}

Since the bootstrap method efficiently works in the single particle models in one-dimension, it is natural to apply it
 to multi-particle systems.
However, we will show that some problems arise when the bootstrap method is applied to (indistinguishable) identical particle systems.
In order to clarify this issue, we study the following one-dimensional two-particle system,
\begin{align}
	H=\frac{p_1^2}{2}+\frac{p_2^2}{2}+V(x_1,x_2).
	\label{H-2bdy}
\end{align}
Here $x_i$ and $p_i$ are the position and momentum of the $i$-th particle ($i=1,2$).
We impose a condition $V(x_1,x_2)=V(x_2,x_1)$ on the potential.
Then, the energy eigenstate is always symmetric or antisymmetric under the exchange of the two particles
$x_1,p_1 \leftrightarrow x_2,p_2$.
It is also possible that the symmetric and anti-symmetric states degenerate at the same energy level.
Related to this property, this model have three different situations, depending on the statistics of the particles.
\begin{itemize}
	\item Two identical bose particles  $\Rightarrow$ States are symmetric under the two particle exchange.
	\item Two identical fermi particles $\Rightarrow$ States are anti-symmetric  under the two particle exchange.
	\item Two distinct particles $\Rightarrow$ State are symmetric or anti-symmetric under the two particle exchange.
\end{itemize}
Note that we do not consider spins.
The distinct particle case can be regarded as a system with a flavor symmetry.

Here, we argue whether the bootstrap method correctly derive the physical quantities in each of these three situations.
As we have seen in the previous section, in order to specify the desired physical situation in the bootstrap method, we need to impose appropriate constraints on expectation values.
In the case of the identical particles, physical quantities are invariant under the exchange of the two particles $x_1,p_1 \leftrightarrow x_2,p_2$, while they need not be invariant in the case of the distinct particles.
Hence, we impose the constraint
\begin{align} 
\langle O(x_1,x_2,p_1,p_2) \rangle = \langle O(x_2,x_1,p_2,p_1) \rangle
\label{constraint-same}
\end{align}
in the identical particle case, and we do not impose it in the distinct particle case. 

In the case of the identical particles, we need to further distinguish the bosons and fermions.
However, the bootstrap method cannot do it.
This is because the quantities considered in the bootstrap method are only expectation values, which are always invariant under the exchange of the particles as in \eqref{constraint-same}.
If the bootstrap method could handle amplitudes or wave functions, it would be possible to distinguish the bosons and fermions, since their sign are flipped under the particle exchange in the case of the fermions. However, there is currently no known way to treat these quantities in the bootstrap method.
Therefore, the bootstrap method cannot distinguish the bosons and fermions.
Hence, the bootstrap method may be applicable to the identical particles but the predictions would be limited.
On the other hand, the bootstrap method may work properly in the distinct particle case.

\subsubsection{Example 1: Non-interacting harmonic oscillators}
\label{sec-HO-2bdy}
To see the problem of two-particle systems in the bootstrap method concretely, we investigate non-interacting two-harmonic oscillators,
\begin{align}
	H= \sum_{i=1}^2  \left( \frac{1}{2}p_i^2+\frac{1}{2} x_i^2\right).
	\label{H-harmonic-2bdy}
\end{align}
We derive the bounds on $\langle x_1 \rangle$ under the constraint $E =\langle H \rangle$.
Before studying the derivation through the bootstrap method, we show the correct bounds obtained through a different method,
\begin{eqnarray} 
& \text{Two distinct particles:} & 	\langle x_1 \rangle^2 
\le 2(E - \hbar), 
\label{x-HO-distinct-exact}
\\
& \text{Two identical bose particles:} & 	\langle x_1 \rangle^2 
\le E - \hbar,
\label{x-HO-bose-exact}
\\
& \text{Two identical fermi particles:} & 	\langle x_1 \rangle^2 
\le E - 2\hbar .
\label{x-HO-fermi-exact}
\end{eqnarray}
The derivation of this result is shown in Appendix \ref{app-HO-2bdy}.
In this computation, we used the property that the harmonic oscillators are quadratic.
However, since this derivation is limited to quadratic systems, it is desirable to reproduce these results employing the uncertainty relation.
(If the uncertainty relation works, we expect that the bootstrap method, which generalizes the uncertainty relation, may work for more general systems.)

Similar to the derivation \eqref{constraint-HO-x} in the single particle problem, the condition $E=\langle H \rangle$ and the uncertainty relation lead to the inequality,
\begin{align}
 \langle x_1 \rangle^2+ \langle x_2 \rangle^2  
\le 2(E - \hbar).
\label{constraint-HO-x-2bdy}
\end{align}
If the two particles are distinguishable, $ \langle x_1 \rangle$ and $ \langle x_2 \rangle$ are independent, and we obtain the bound
\begin{align}
	\langle x_1 \rangle^2 
   \le 2(E - \hbar).
   \label{x-HO-distinct}
\end{align}
This reproduces \eqref{x-HO-distinct-exact}, and thus the uncertainty relation (and the bootstrap method) may work for the distinct particles. 

If the two particles are identical and not distinguishable, 
the relation $ \langle x_1 \rangle= \langle x_2 \rangle$ is satisfied through \eqref{constraint-same}, and the inequality \eqref{constraint-HO-x-2bdy} becomes
   \begin{align}
	\langle x_1 \rangle^2 
   \le E - \hbar.
   \label{x-HO-identical}
\end{align}
No further restrictions can be imposed from the uncertainty relation.
This result is consistent with the boson \eqref{x-HO-bose-exact}, but not with the fermion \eqref{x-HO-fermi-exact}.
Clearly, the bound \eqref{x-HO-identical} is weaker than the desired bound \eqref{x-HO-fermi-exact} for the fermions.
This means that the condition obtained from the uncertainty relation is not strong enough.
This is due to the fact that, as mentioned earlier, the uncertainty relation handles only expectation values, which do not distinguish bosons and fermions.

Similar problems must appear in more general models in the bootstrap methods too.
The related issues would also arise in the derivation of the energy eigenstates.
We will consider this problem in the next example.

\subsubsection{Example 2: Yang-Mills quantum mechanics}
\label{sec-YMQM}

\begin{figure}
	\begin{center}
		\includegraphics[scale=0.6]{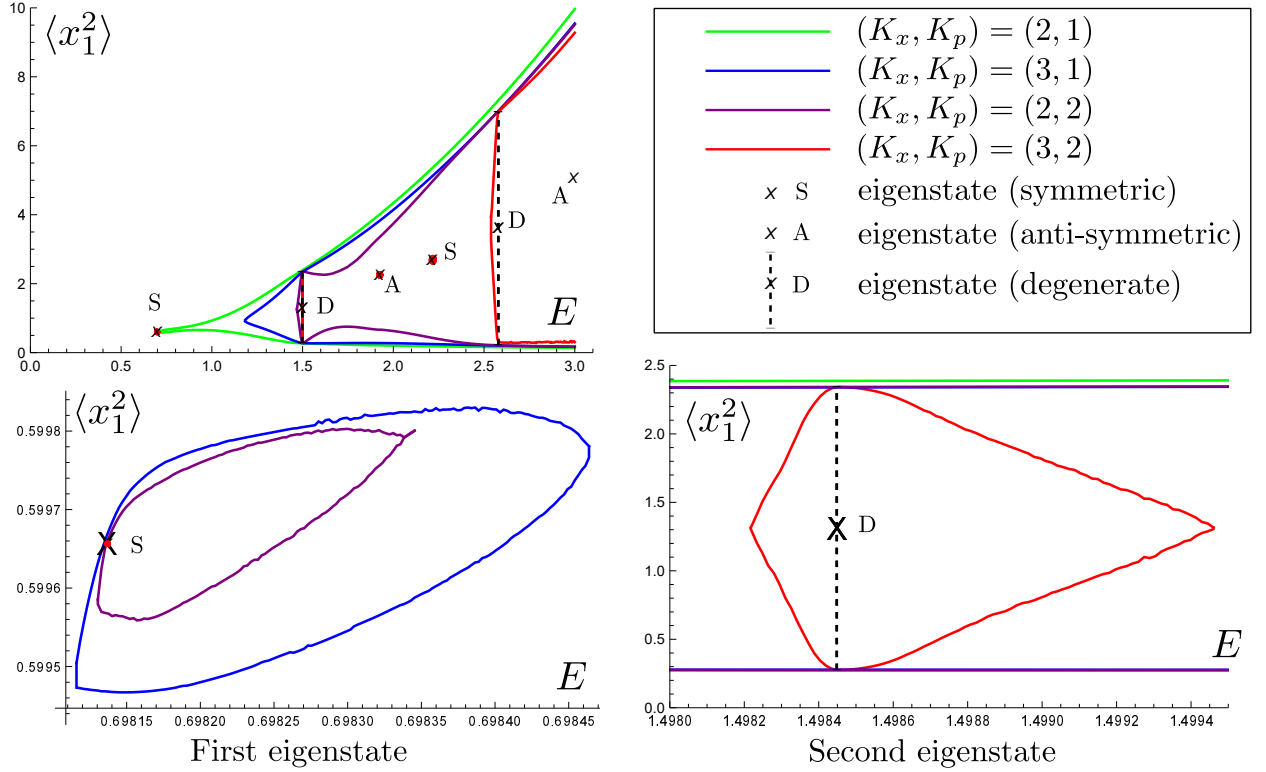}
	\end{center}
	\caption{
		Energy eigenstates for the distinct two-particles in the YMQM \eqref{H-YMQM}.
		``$\times$" denotes the eigenstates obtained by solving the Schr$\ddot{\rm o}$dinger equation numerically.
		The lower left and the lower right panels are the neighborhood of the first and second eigenstates in the upper left panel, respectively.
		The eigenstates are symmetric or antisymmetric with respect to the exchanges of the particles.
		In particular, when these two states degenerate, $\langle x^2_1 \rangle$ can take various values within a certain range due to their superpositions.
		This range is indicated by the vertical dashed lines in the figures.
		It can be seen that the bootstrap method reproduces these eigenstates including the superpositions as $K_x$ and $K_p$ increase. 
		Particularly, the first eigenstate at $(K_x, K_p)=(3,2)$ (the red ``dot") is almost point-like.
		However, the bootstrap method cannot indicate whether these eigenstates are symmetric or antisymmetric.	}
	\label{Fig-YMQM-eigen}
\end{figure}

\begin{figure}
	\begin{center}
		\includegraphics[scale=0.6]{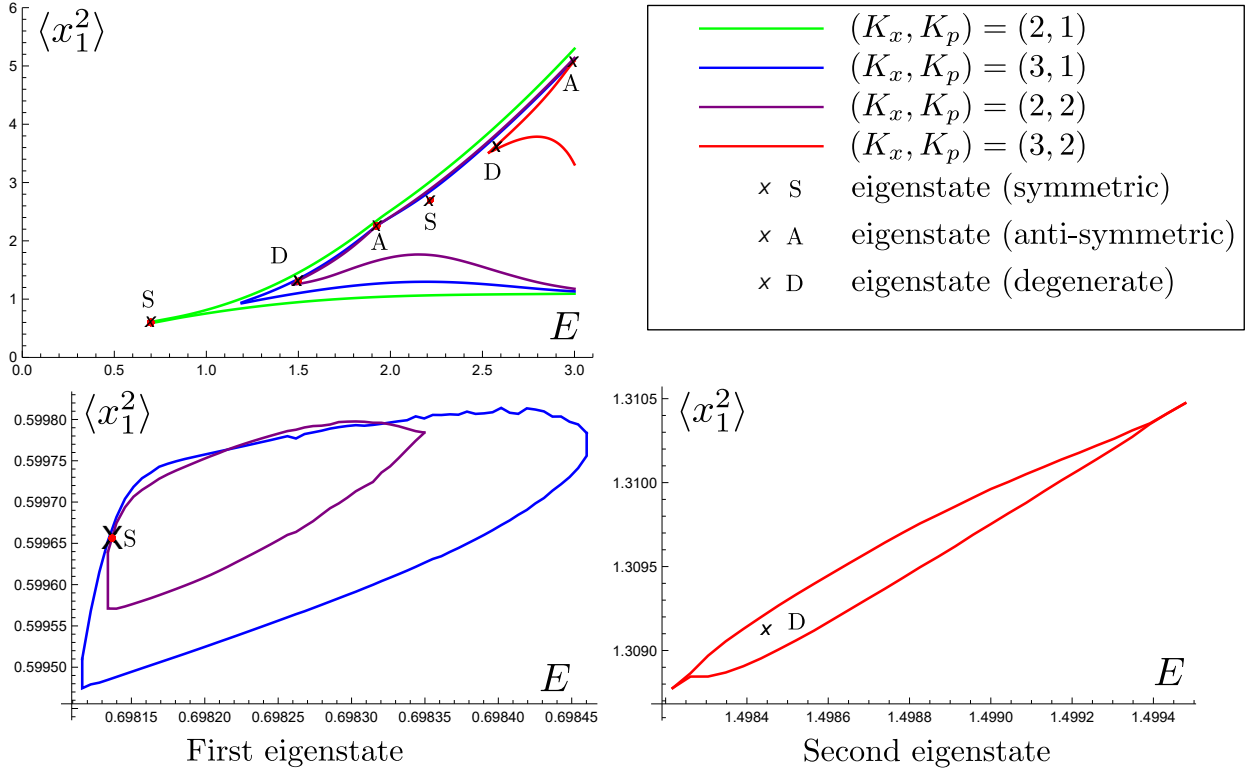}
	\end{center}
	\caption{
		Energy eigenstates for the identical two-particles in the YMQM \eqref{H-YMQM} through the numerical bootstrap method.
		``$\times$" denotes the eigenstates obtained by solving the Schr$\ddot{\rm o}$dinger equation numerically.
		The lower left and the lower right panels are the neighborhood of the first and second eigenstates in the upper left panel, respectively.
		The eigenstates are symmetric or antisymmetric with respect to the particle exchanges, and we take either of them depending on whether the particles are the bosons or the fermions. 
		At the degenerate states, the bosons and fermions have the same values $\langle x^2_1 \rangle$.
		Different from the distinct particle case, their superposition is not allowed, since we take only one state.
		It can be seen that the bootstrap method reproduces these states as $K_x$ and $K_p$ increase. 
		The first eigenstate at $(K_x, K_p)=(3,2)$ is again almost point-like.
		However, the bootstrap method cannot indicate whether the states are symmetric or antisymmetric.	
	}
	\label{Fig-YMQM-eigen-sym}
\end{figure}

As the second example, we study so called ``Yang-Mills quantum mechanics" (YMQM), which is not free and known to show chaos \cite{Matinyan:1981dj, Savvidy:1982jk, Akutagawa:2020qbj},
\begin{align}
	H=\frac{p_1^2}{2}+\frac{p_2^2}{2}+x_1^2 x_2^2.
	\label{H-YMQM}
\end{align}
Numerical computation of the Schr$\ddot{\rm o}$dinger equation yields symmetric and antisymmetric eigenfunctions.
There are also cases where both degenerate.
In order to compute the observables such as $\langle x_1^2 \rangle$, we need to select the eigenfunctions depending on the three situations: the distinct particles, the bosonic identical particles and the fermionic identical particles.
The spectra $(E, \langle x_1^2 \rangle) $ in these cases are plotted in Fig.~\ref{Fig-YMQM-eigen} for the distinct particles and Fig.~\ref{Fig-YMQM-eigen-sym} for the identical particles.
Note that, in the case of the distinct particles, when two eigenstates degenerate, 
superpositions of these two are allowed, and the expectation values of $\langle x_1^2 \rangle $ at this energy level take various values within a certain range.
The black vertical dashed lines in Fig.~\ref{Fig-YMQM-eigen} indicate this range.
On the other hand, in the case of the identical particles, we need to exclude the anti-symmetric states or the symmetric states according to whether the particles are bosons or fermions, and the degeneracy does not occur.

Now, let us see if the bootstrap method can reproduce these eigenstates.
We take the seed operator
\begin{align}
	\tilde{O}= \sum_{k,l=0}^{K_x} \sum_{m,n=0}^{K_p} b_{klmn} x_1^k x_2^l p_1^m  p_2^n,
	\label{Otilde-xpyq}
\end{align}
and construct the bootstrap matrix from it.
Then, we solve the optimization problem \eqref{constraint-eigen}, which is linear if we fix $E$.
We also impose the constraints \eqref{constraint-same} for the particle exchanges in the  identical particle case.
The numerical results are illustrated in Fig.~\ref{Fig-YMQM-eigen} (the distinct particles) and Fig.~\ref{Fig-YMQM-eigen-sym} (the identical particles).\footnote{
We solve this problem by using the Mathematica package ``SemidefiniteOptimization" with the option ``Mosek" and ``CSDP".
In our numerical analysis, sometimes unnatural line-like regions were observed. In these regions, the bootstrap matrices have much larger negative eigenvalues, which mean the condition ${\mathcal M}
\succeq 0$ is not satisfied and the data would be not reliable.
In our figures, we have removed these regions.
}
They show that, for both identical and distinct particles, the range of the possible values of $\langle x^2_1 \rangle$ tends to approach to the eigenstates as $K_x$ and $K_p$ increase.
In the distinct particle case, the numerical bootstrap method also reproduces the range of $\langle x_1^2 \rangle$ due to the degeneracy.

However, as we have discussed, it is not possible to determine whether the obtained eigenstates are for the bosons or for the fermions through the bootstrap method.
(We can only say that there are states corresponding to the bosons and the fermions at the degenerate states from the results for the distinct particles.)
Therefore, the bootstrap method has only limited predictive power for the identical particle case.

Note that the number of the independent variables in the bootstrap matrix ${\mathcal M}$ in the YMQM increases as $K_x$ and $K_p$ increases, ant it is more than 400 at $(K_x,K_p)=(3,2)$ even after we impose the constraints in \eqref{constraint-eigen}.
In contrast, in the case of the one-dimensional anharmonic oscillator argued in Sec.~\ref{sec-eigen-example}, the independent variables are only two: $\langle x \rangle$ and $\langle x^2 \rangle$, and it does not change by $(K_x,K_p)$ because of the strong constraint \eqref{AHO-recurrence}.
Thus, the numerical bootstrap analysis in the YMQM case is qualitatively different from the anharmonic oscillator case, and our results show that the method works even in such a situation.

\subsubsection{Bootstrapping other states with $E=\langle H \rangle$ in YMQM}
\label{sec-YMQM-other}

\begin{figure}
	\begin{center}
		\includegraphics[scale=0.5]{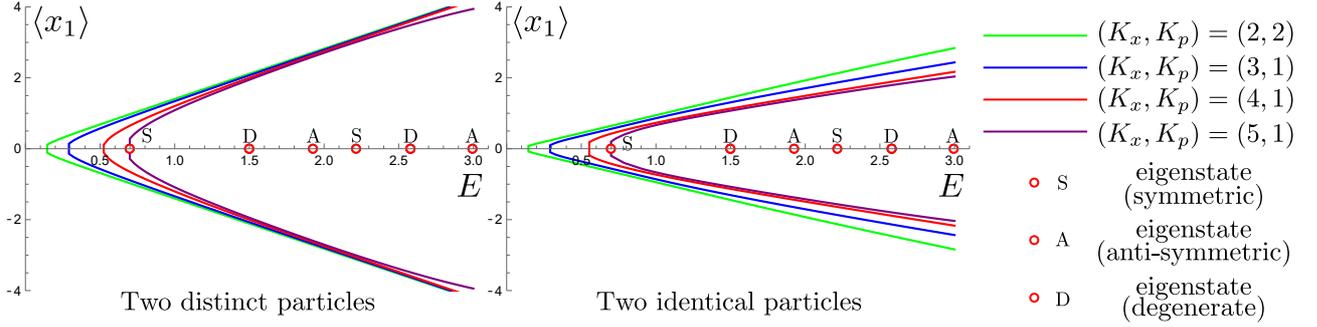}
	\end{center}
	\caption{
		The possible ranges of $\langle x_1 \rangle$ in the YMQM for general mixed states with $E=\langle H \rangle$.
		The left panel is for the distinct particles and the right panel is for the identical particles in which the constraint \eqref{constraint-same} is imposed.
		Thus, the region in the right panel is smaller.
	}
	\label{Fig-YMQM-x1}
\end{figure}

\begin{figure}
	\begin{center}
		\includegraphics[scale=0.5]{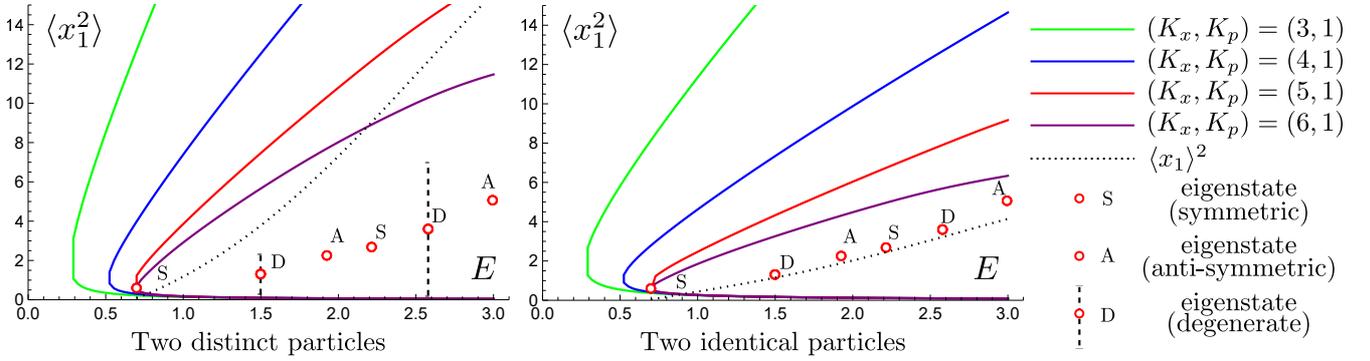}
	\end{center}
	\caption{
		The possible ranges of $\langle x^2_1 \rangle$ in the YMQM for general mixed states with $E=\langle H \rangle$.
		The left panel is for the distinct particles and the right panel is for the identical particles.
		It seems that the bootstrap results have not converged yet at $(K_x,K_p)=(6,1)$.
		We also plot $\langle x_1 \rangle^2$ at $(K_x,K_p)=(5,1)$, which is obtained from the data in Fig.~\ref{Fig-YMQM-x1}.
		We see that $\langle x^2_1 \rangle > \langle x_1 \rangle^2$ is not satisfied for larger $E$ in the distinct particle case. It is an evidence that the analysis for $ \langle x_1 \rangle$ have not been converged yet. Taking larger $K_x$ and $K_p$ might improve it. 
	}
	\label{Fig-YMQM-x2}
\end{figure}

As we have studied in the one-dimensional quantum mechanics, the bounds on the possible values of observables for general mixed states with energy $E=\langle H \rangle_\rho$ in the YMQM will be derived through the bootstrap method.
We solve the optimization problem \eqref{constraint-general-E} with the Hamiltonian \eqref{H-YMQM} by using the bootstrap matrix constructed from the operator \eqref{Otilde-xpyq}.
The results for $\langle x_1 \rangle$ and $\langle x_1^2 \rangle$ are illustrated in Fig.~\ref{Fig-YMQM-x1} and \ref{Fig-YMQM-x2}, respectively.\footnote{
	In our analysis in this subsection, we use the Mathematica package ``SemidefiniteOptimization" with the option ``DSDP".
	Note that $\langle x_1 \rangle$ for the energy eigenstates are always zero even superposing the degenerate states in the distinct particle case.
	}
It seems that  $\langle x_1^2 \rangle$ have not converged yet.
 $\langle x_1 \rangle$ in the distinct particle case also have not converged yet, since they do not satisfy $\langle x_1 \rangle^2 <  \langle x_1^2 \rangle$ for larger $E$ as shown in Fig.~\ref{Fig-YMQM-x2}.
(Another possibility is that the obtained $\langle x_1^2 \rangle$ in this region is numerically wrong.)  
They might have converged for smaller $E$. $\langle x_1 \rangle$ in the identical particle case might have converged too.
In order to obtain the convergent results, we need to perform numerical analysis for larger $K_x$ and $K_p$, but, due to our limited computational resources, we leave it as a future works.

\begin{figure}
			\begin{center}
				\includegraphics[scale=0.5]{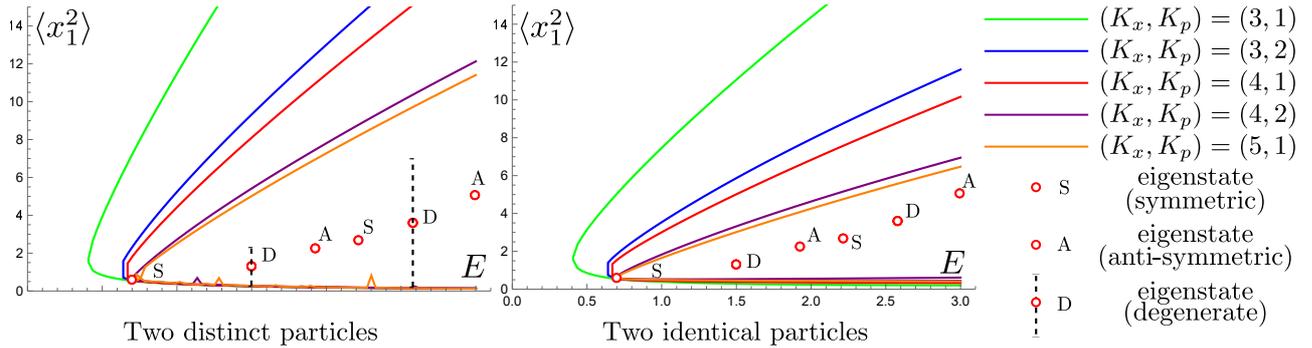}
			\end{center}
			\caption{
				The possible ranges of $\langle x^2_1 \rangle$ in the YMQM for stationary states with $E=\langle H \rangle$.
				}
			\label{Fig-YMQM-static-x2}
		\end{figure}

We also investigate stationary states with energy $E=\langle H \rangle_{\rho_{\rm st}}$ by solving the constraint \eqref{constraint-statonary-E}.
The results for $\langle x_1^2 \rangle_{\rho_{\rm st}}$ are shown in Fig.~\ref{Fig-YMQM-static-x2}.\footnote{
		To improve the numerical analysis for stationary states, we impose the parity condition $\langle x^k_1 x^l_2 p^m_1 p^n_2  \rangle =0 $ if either $k+m$ or $l+n$ is odd.
		Note that, in the single particle case, the bounds for the stationary states are derived through the straight lines connecting the spectrum of the particle as shown in Fig.~\ref{Fig-static}. This method does not work in the YMQM, since the spectrum for large energy is not known. (In the single particle case, the spectrum for large energy can be obtained through the WKB approximation. However, the WKB approximation does not work in the YMQM.)
		}

In all of the bootstrap results in this subsection, the regions in the identical particle case are always smaller than those in the distinct particle case because of the additional constraint \eqref{constraint-same}.
In the identical particle case, the actual allowed regions must be smaller than the obtained regions in our analysis, similar to the harmonic oscillator case discussed in Sec.~\ref{sec-HO-2bdy}.
Particularly, the region for the fermions should start from the ground state, but the bootstrap method cannot show it at all.

Note that the YMQM \eqref{H-YMQM} has flat directions along the line $x_1=0$ and $x_2=0$.
Thus, $x_1$ or $x_2$ can take arbitrary large values even at zero energy in classical mechanics.
On the other hand, in quantum mechanics, the probability of taking such a large value $x_i$ is expected to be small due to the uncertainty relation \cite{SIMON1983209}.
Our results for $\langle x_1 \rangle$ and $\langle x^2_1 \rangle$ explicitly 
support this prediction.
\\

In this section, we have discussed the application of the bootstrap method to two-particle systems in one-dimension.
In the case of the identical particles, the bootstrap method has only limited predictive power.
On the other hand, it properly works when the two particles are distinct.
In this case, $x_1$ and $x_2$ can be interpreted as ``the coordinates of one particle in two dimensions" instead of ``the coordinates of two particles in one dimension."
Therefore, the bootstrap method works even in two dimension system.
By increasing the number of degrees of freedom in this way, we expect that it will also work for multi-particles in one or higher dimensions.

\section{Bootstrapping Thermal Equilibrium States}
\label{sec-thermal}

In the previous section, we have presented that the bootstrap method also works for multi-particle systems, although the predictive power is limited in the identical particle cases.
An interesting question is whether the bootstrap method can predict physical quantities in thermal equilibrium in quantum many-body systems.
In this section, we discuss this problem.
To make the problem concrete, we consider an $N$-particle system ($N \gg 1 $) with Hamiltonian
\begin{align} 
	H= \frac{1}{2} \sum_{i=1}^N p_i^2 + U(x_1, x_2, \cdots x_N),
	\label{H-many}
\end{align}
where $x_i$ is the position of the $i$-th particle and $p_i$ is its conjugate momentum.
$U(x_1, x_2, \cdots x_N)$ is a potential and we assume that it is symmetric with respect to the particle exchanges.
(The large-$N$ matrix models studied in \cite{Han:2020bkb} are examples of this model.)

Obviously, if we consider identical particles in the model \eqref{H-many}, the issue of identical particles would occur.
In addition, we find other problems that, in principle, it is difficult to handle temperature and entropy in the bootstrap method.
In this section, we first introduce the temperature problem in Sec.~\ref{sec-canonical}, and after that we discuss entropy and identical particle problems in Sec.~\ref{sec-micro}.
However, if the system is integrable, the bootstrap method may evade the identical particle problem. We will argue it in Sec.~\ref{sec-free}.

\subsection{Difficulties in bootstrapping canonical ensemble}
\label{sec-canonical}

In this section, we try to apply the bootstrap method to the model \eqref{H-many} in thermal equilibrium with temperature $T$.
Such a state is described by the stationary mixed state \eqref{mixed-static} with the Boltzmann factor $c_n=\exp(-\beta E_n)/Z$, and 
 the expectation value of an operator $O$ is given by
\begin{align} 
	\langle O \rangle_{\beta}:= \frac{1}{Z} \sum_n e^{-\beta E_n} \langle n |  O | n \rangle , 
	\qquad Z:=\sum_n \langle n | e^{-\beta H}  | n \rangle. 
	\label{canonical}
\end{align}
Here $\beta:=1/T$ and we have taken the Boltzmann constant 1.
Typically, we take $O$ as an averaged macroscopic quantity, for example,
\begin{align}
	\overline{x^mp^n} :=  \frac{1}{N} \sum_{i=1}^N  x^m_i p^n_i.
\label{x-average}
\end{align}

In order to evaluate physical quantities in this thermal equilibrium state using the bootstrap method, constraints that specify this state should be imposed on quantities $\langle O \rangle_{\beta}$.
Since thermal equilibrium state is a kind of stationary mixed states, the constraint \eqref{HO=0} should be imposed.
Then, we need to find additional constraints such that we distinguish the thermal equilibrium state from general stationary mixed states.

We notice that the thermal equilibrium state \eqref{canonical} satisfies the condition,
\begin{align}
	\frac{\partial}{\partial \beta}	\langle O \rangle_{\beta}=
	\langle H \rangle_{\beta} \langle O \rangle_{\beta} - \langle HO \rangle_{\beta}.
	\label{beta-O}
\end{align}
(We can also find similar equations for the higher-order derivative of $\beta$.)
However, this condition is a differential equation with respect to $\beta$, which is not useful at all in the bootstrap method.\footnote{
	Since the relation $\eqref{beta-O}$ is a differential equation, if we know the expectation value $\langle O \rangle_{\beta}$ at a certain temperature, we may use it to evaluate the physical quantity at a slightly different temperature $\beta+\Delta \beta$.
	However, the bootstrap method is not so useful in combination with differential equations, because it is a method to test whether the expectation values $\langle O \rangle_{\beta}$ is consistent with the given constraints.
}
On the other hand, this relation gives us indirectly useful information. 
That is, when the system size is large, the expectation values of averaged quantities such as \eqref{x-average} should be factorized,
\begin{align}
	\langle O_1 O_2 \rangle_{\beta}
	=\langle O_1 \rangle_{\beta} \langle O_2 \rangle_{\beta} + O(1/N).
	\label{constraint-factorization}
\end{align}
Since $\langle O_1 \rangle$ and $\langle O_2 \rangle$ would be $O(1)$ quantities, we may ignore the second term on the right-hand side. 
Thus, if this relation does not hold, the order of $N$ does not match on the right and left-hand sides of the equation \eqref{beta-O}.
Therefore, such factorization must occur in the thermal equilibrium state \eqref{canonical}.
(Conversely, factorization does not need to occur in general stationary mixed states.)
In fact, in statistical mechanics, such a relation is naturally expected from the central limit theorem.
In addition, in large-$N$ gauge theories, such a relation is also expected as large-$N$ factorizations\footnote{Large-$N$ factorizations do not need to occur in arbitrary states in large-$N$ gauge theories.}.
So, the factorization \eqref{constraint-factorization} is one of the conditions that distinguish the
thermal equilibrium state from general stationary mixed states.

Note that, once we impose the factorization condition \eqref{constraint-factorization}, the commutator relation \eqref{HO=0} seems trivial. 
However, it provides important relations at $O(1/N)$, and we pick them up when we use the bootstrap method. Similar things happen in the relation \eqref{beta-O} too, although we will not use \eqref{beta-O} in our bootstrap analysis.

However, we could not find any other useful conditions, which characterize the thermal equilibrium state.
In particular, the factorization condition \eqref{constraint-factorization} has no information on temperature.
This implies that temperature cannot be handled in the bootstrap method.
In other words, it is difficult for the bootstrap method to predict the expectation values of observables at a given temperature.
(The only exception is the ground state corresponding to zero-temperature.)

Related to the difficulty of handling temperature in the bootstrap method, chemical potentials cannot be handled either.
These results suggest that the bootstrap method has difficulty in dealing with (grand) canonical ensembles specified by temperature and chemical potentials.

\subsection{Bootstrapping micro-canonical ensemble}
\label{sec-micro}

We have seen that it is difficult to investigate the temperature dependence in thermal equilibrium using the bootstrap method.
However, the bootstrap method allows us to specify energy as $E = \langle H \rangle$.
Therefore, there is a possibility to evaluate physical quantities in thermal equilibrium as a micro-canonical ensemble.
Specifically, we can impose the following constraints,
\begin{align}
	{\mathcal M}
	\succeq 0, \quad E = \langle H \rangle,  \quad  	\langle [ H,O] \rangle=0, \quad \langle O_1 O_2 \rangle	=\langle O_1 \rangle \langle O_2 \rangle,
	\label{constraint-micro-canonical}
\end{align}
and investigate the range of possible values of $\langle Q \rangle$.
If the obtained range is sufficiently narrow, the value $\langle Q \rangle$ may correspond to that of the micro-canonical ensemble at the given energy $E$.

However, recall that the bootstrap method cannot distinguish the two bosons and two fermions in the two particle systems.
For the $N$-particle system, there are much more possibilities on the particle species such as multiple species of bosons and fermions.
Then, it may be difficult to obtain convergent results through the bootstrap method which cannot distinguish them.

In addition, even if we obtain some reliable and convergent results, the bootstrap method cannot give us the entropy.
This is because the bootstrap method only tells us whether the value of a physical quantity is consistent with quantum mechanics or not, and it does not tell us the degeneracy.

Note that, if we employ a quantum field for describing a single species particles, the issue of the convergence arising through the particle statistics may be evaded. 
Similarly, this issue may not exist in lattice systems.
It is valuable to investigate the bootstrap method in these models, and we leave this challenge to future work.

\subsection{Bootstrapping integrable systems in micro-canonical ensembles}
\label{sec-free}

So far, we have discussed the difficulties on the bootstrap method for thermal equilibrium states.
It can be applied to micro-canonical ensembles only, and, even in this case, the results may not converge and entropy cannot be derived either. 
However, we argue that the issue of the convergence may be resolved in integrable systems.

In integrable systems, there are numerous conserved charges, 
and thermal equilibrium states are specified by these charges or their conjugate chemical potentials. 
(The grand canonical ensemble characterized by these numerous chemical potentials is called ``generalized Gibbs ensemble" (GGE) \cite{PhysRevLett.98.050405}, and are actively studied recently. See review articles \cite{Polkovnikov:2010yn, DAlessio:2015qtq}.)

In the micro-canonical ensemble, the numerous conserved charges may fix the physical quantities in the bootstrap method.
In Appendix \ref{app-GGE}, we study two integrable systems: non-interacting $N$-harmonic oscillators and non-interacting $N$-anharmonic oscillators, and find that the correct results are obtained in these models through the bootstrap method (see Fig.~\ref{Fig-thermal-free}).

\subsubsection{Thermometer?}
\label{sec-meter}

We have discussed that the bootstrap method can reproduce physical quantities at thermal equilibrium in the integrable systems.
However, as discussed in the previous section, temperature and entropy cannot be evaluated through the bootstrap method.

One possibility to obtain temperature is to introduce a ``thermometer".
We prepare a system whose spectrum is well known, for example, harmonic oscillator, as a thermometer, and turn on weak interactions between this thermometer system and the target system, which we want to investigate.
Then, from the spectrum of the thermometer system, we might be able to read the temperature of the entire system, and we might obtain the temperature dependence of the target system.
However, it is unclear whether the bootstrap method works in such a interacting system, and we leave this as a future problem.

\section{Discussions}
\label{sec-discussion}

We have presented that the bootstrap method can be used to obtain the bounds of possible expectation values of various physical quantities under the constraint $E=\langle H \rangle$. 
The bootstrap method can be regarded as a generalization of the uncertainty relation, and the bounds of such physical quantities are determined as a consequence of these uncertainty relations.
We have also argued that the bounds of the possible values are further restricted through the additional constraints $\langle [H,O] \rangle=0$ for the stationary states \eqref{constraint-statonary-E} and $\langle [H,O] \rangle=0$ and $\langle HO \rangle= E \langle O \rangle$ for the energy eigenstates \eqref{constraint-eigen}.
In this way, the difference between these three states in quantum mechanics is described by the difference of the constraints on expectation values in the bootstrap method.

These properties may reveal novel aspects of quantum mechanics.
Particularly, our results indicate that the energy eigenstates may be determined through the uncertainty relations in a broad sense.
It may be valuable to pursue this question and understand these algebraic structures of quantum mechanics further.

On the other hand, we found that there are no suitable constraints that describe the difference between the identical bosons and fermions.
It is also difficult to describe the thermal equilibrium states in canonical ensemble because there are no useful constraints which specify the temperatures.
These difficulties may be a sort of no go theorems, and it may be valuable to investigate these issues further.
There might be some profound reasons why the bootstrap method does not work in these situations.

Another interesting direction is applying the bootstrap problem to lattice models and quantum field theories \cite{Lawrence:2021msm,Anderson:2016rcw, Lin:2020mme, Kazakov:2021lel, Kazakov:2022xuh, Cho:2022lcj}.
Especially, the issue of the identical particles might be evaded in these systems.
We leave this as a future problem.

\paragraph{Acknowledgements}
The author would like to thank Takehiro Azuma, Masafumi Fukuma, Koji Hashimoto, Satoru Odake, Junji Suzuki and Asato Tsuchiya for valuable discussions and comments.
The author would also like to thank Yu Nakayama for valuable discussions and pointing out some issues in his numerical analysis on the YMQM.  
The author is especially grateful to Yu Aikawa and Kota Yoshimura for very helpful correspondence and collaboration at an early stage of this project. 
The author would like to thank participants of the YITP workshop ``Strings and Fields 2021" (YITP-W-21-04, 23 August to 27 August 2021),  ``Strings and Fields 2022" (YITP-W-22-09, 19 August to 23 August 2022)  and ``Thermal Quantum Field Theory and Their Applications 2021" (KEK, 30 August to 1 September 2021)  for stimulating discussions where part of this work was presented.
Part of numerical computation in this work was carried out at the Yukawa Institute Computer Facility.
The work of T.~M. is supported in part by Grant-in-Aid for Scientific Research C (No. 20K03946) from JSPS.

\appendix

\section{Analytic Results on the Bounds}
\label{app-bounds}
In this appendix, we explain details of some analytical results on the problems of finding the bounds on the expectation values.

\subsection{States saturating the bound \eqref{x-bound-HO}}
\label{app-HO-x}

We derive the states that saturate the inequality \eqref{x-bound-HO} in the harmonic oscillator \eqref{H-harmonic}.
This inequality is saturated when all the inequalities in Eq.~\eqref{constraint-HO-x} are saturated.
Therefore $\Delta x^2 = \Delta p^2 = \hbar/2 $ must be satisfied.
This is the relation satisfied by coherent states.
From this, we find the states $\exp \left( \mp ipx_* (E) /\hbar \right) |0 \rangle $,
where $ |0 \rangle $ is the ground state of the harmonic oscillator and $x_* (E)$ is defined in \eqref{x-bound-HO}.
($ |0 \rangle $ is translated by $\pm x_* (E)$ in this state.)
Actually, we can easily show that these states satisfy $E=\langle H \rangle $ and saturate the bounds \eqref{x-bound-HO} as,
\begin{align} 
	\langle 0| e^{ipx_* /\hbar} H  e^{-ipx_* /\hbar}  |0 \rangle = 
	\langle 0| H    |0 \rangle + \frac{1}{2}\left( x_*(E) \right)^2 = E, \quad
	\langle 0| e^{ipx_* /\hbar} x  e^{-ipx_* /\hbar}  |0 \rangle = x_*(E)
	.
	\label{HO-coherent}
\end{align}
Here we have used $ \langle 0| x |0 \rangle =0$ and $\langle 0| H    |0 \rangle=\hbar/2$.

Note that it is easy to show that any coherent states with energy $E=\langle H \rangle $
satisfy
\begin{align}
	  \langle p \rangle^2 +  \langle  x \rangle^2 =2(E - \hbar/2).
\end{align}
Since the Heisenberg equations for $(\langle x(t) \rangle, \langle p(t) \rangle)$ is equivalent to the classical equation of motion in the harmonic oscillator, the coherent states always pass the points $(\langle x \rangle, \langle p \rangle) =  ( \pm x_*(E),0)$ through the time evolution.
Therefore, the coherent states always saturate the bounds.

\subsection{States saturating the bound \eqref{x2-bound-HO}}
\label{app-HO-x2}

We derive the states that saturate the inequality \eqref{x2-bound-HO} in the harmonic oscillator \eqref{H-harmonic}.
This inequality is saturated when $\langle x^2 \rangle  \langle p^2 \rangle = \hbar^2/4$ is satisfied, and we know that gaussian wave packets with $\langle x \rangle  = \langle p \rangle = 0$ satisfy it.
Indeed, if we take the deviation of the gaussian wave packets as
\begin{align} 
	\langle x^2 \rangle  = E+ \sqrt{E^2 - \hbar^2/4}, 
\end{align}
$\langle x^2 \rangle$ saturates the upper bound of \eqref{x2-bound-HO}.
Then this gaussian wave packet satisfies $\langle p^2 \rangle  = E- \sqrt{E^2 - \hbar^2/4}$, and we obtain $\langle H \rangle =E$.
Similarly, if we take $\langle x^2 \rangle  = E- \sqrt{E^2 - \hbar^2/4}$, it saturates the lower bound. 

The coherent states discussed in Appendix \ref{app-HO-x} and the gaussian wave packets are fundamental states in quantum mechanics.
It is an interesting conclusion that these states have the properties of maximizing (minimizing) $\langle x \rangle$ and $\langle x^2 \rangle$ in the harmonic oscillator.

\subsection{Bounds on $\langle p \rangle$ in general $V(x)$}
\label{app-p}

We derive the upper and lower bounds on $\langle p \rangle$ for general non-relativistic quantum mechanical systems,
\begin{align} 
	H=\frac{1}{2}p^2+V(x).
	\label{H-general-app}
\end{align}
We will show that the answer is given as
\begin{align} 
	-p_* \le  \langle p \rangle \le p_*, \qquad p_*(E):=\sqrt{2(E-E_0)},
	\label{p-bound}
\end{align}
where $E_0$ is the energy of the ground state.
We prove it by contradiction.
Suppose a state $ | \alpha \rangle$ satisfies $E= \langle \alpha |H| \alpha \rangle   $ and 
$\tilde{p}= \langle \alpha |p| \alpha \rangle   $, where $\tilde{p} >  p_*(E) $ and violates the bound \eqref{p-bound}.
Then the state $ e^{-i \tilde{p} x/ \hbar} | \alpha \rangle$ satisfies
\begin{align} 
	\langle \alpha | e^{i \tilde{p} x / \hbar} H e^{-i \tilde{p} x / \hbar} | \alpha \rangle=
	\langle \alpha |  H  | \alpha \rangle + \frac{1}{2} \tilde{p}^2 
	- \tilde{p} \langle \alpha |  p  | \alpha \rangle
	=
	E - \frac{1}{2} \tilde{p}^2  < E_0.
\end{align}
Thus, the energy is lower than the ground state, and the state $ | \alpha \rangle$ is inconsistent. 
Hence, there is no state which satisfies $\tilde{p} >  p_*(E) $ and the bound \eqref{p-bound} is proved.

Besides, we can easily show that the state $ e^{ \pm i p_* x/ \hbar} | 0 \rangle$ saturates the inequalities in \eqref{p-bound}, where $ | 0 \rangle$ is the ground state.
The proof is similar to \eqref{HO-coherent} but we need to use $ \langle 0| p | 0 \rangle  =0$, which can be shown by using the relation \eqref{HO=0} with $O=x$.

The result \eqref{p-bound} indicates that $E-E_0$ yields the maximum value of  $ |\langle  p  \rangle |$.
This may be reasonable, since $ |\langle  p  \rangle| >0$ always causes an excitation from the ground state.

Note that $p$ is quadratic in the Hamiltonian \eqref{H-general-app}, and it is crucial in the above derivation of the bounds.
Therefore, it seems difficult to apply this method to obtain the bounds on $ \langle x \rangle $ in the Hamiltonian \eqref{H-general-app}.

\subsection{Bounds on $\langle x_1 \rangle$ in non-interacting two-harmonic oscillators}
\label{app-HO-2bdy}

We show the derivation of the bounds \eqref{x-HO-distinct-exact}, \eqref{x-HO-bose-exact} and \eqref{x-HO-fermi-exact} on $\langle x_1 \rangle$ in the non-interacting two-harmonic oscillators,
\begin{align}
	H= \sum_{i=1}^2~ \frac{1}{2}p_i^2+\frac{1}{2} x_i^2.
	\label{H-harmonic-2bdy-app}
\end{align}
Since $x_1$ is quadratic in this Hamiltonian, we can apply the method used in Appendix \ref{app-p}.

The ground states and the ground energies of this system is given by
\begin{eqnarray} 
	& \text{Two distinct particles:} &  | 0 \rangle_2:=	| 0,0 \rangle, \quad E_0= \hbar, 
	\label{0-HO-distinct-exact}
	\\
	& \text{Two identical bose particles:} & 	| 0 \rangle_B:=| 0,0 \rangle, \quad E_0= \hbar, 
	\label{0-HO-bose-exact}
	\\
	& \text{Two identical fermi particles:} & | 0 \rangle_F:= \frac{1}{\sqrt{2}} \left( | 1,0 \rangle -| 0,1 \rangle
	\right), \quad E_0= 2\hbar,
			\label{0-HO-fermi-exact}
	\end{eqnarray}
	where $| m,n \rangle:= (a_1^\dagger)^m (a_2^\dagger)^n | 0,0 \rangle $.
	Then, by translating these ground states by $U(x_*):= e^{-ip_1 x_*}$, we may obtain the states which provide the maximum value of $|\langle x_1 \rangle |$ as in Appendix \ref{app-p}.
	However, this translation operator is not symmetric under the particle exchange.
	Hence, it cannot be used for the identical particles, and we modify it as $e^{-i(p_1+ p_2) x_*}$.
	Here $x_*$ should be determined to satisfy $E=\langle 0| U(x_*)^\dagger H U(x_*) |0 \rangle$ for each ground state, and we obtain
	\begin{eqnarray} 
	& \text{Two distinct particles:} &  E=\langle 0 | H | 0 \rangle_2 + \frac{1}{2}x_*^2 
	\quad \Rightarrow \quad x_*^2=2(E-\hbar), 
	\\
	& \text{Two identical bose particles:} & E=\langle 0 | H | 0 \rangle_B + x_*^2 
	\quad \Rightarrow \quad x_*^2=E-\hbar,
	\\
	& \text{Two identical fermi particles:} & E=\langle 0 | H | 0 \rangle_F + x_*^2 
	\quad \Rightarrow \quad x_*^2=E-2\hbar.
	\end{eqnarray}
	Similar to the proof in Appendix \ref{app-p}, we can show that, if $|\langle x_1 \rangle|$ exceeds $x_*$, it causes a contradiction and such a state is not allowed.
	Thus, $x_*$ provides the maximum bound.

\section{Bootstrapping free particles in micro-canonical ensemble}
\label{app-GGE}

In Sec.~\ref{sec-free}, we have argued that the numerical bootstrap method may determine observables in integrable systems in thermal equilibrium.
In this appendix, we demonstrate it in two models: non-interacting $N$-harmonic oscillators and non-interacting $N$-anharmonic oscillators.

\subsection{Example 1: Non-interacting $N$-harmonic oscillators}
\label{sec-GGE-HO}

We investigate the non-interacting $N$-harmonic oscillators in one-dimension,
\begin{align} 
H= \hbar \sum_{i=1}^N \left( a^\dagger_i a_i + \frac{1}{2} \right), \qquad a_i=\frac{1}{\sqrt{2\hbar}} \left(x_i + i p_i\right).
\label{H-HO-N}
\end{align}
This system has the infinite number of the conserved charges,
\begin{align} 
	R_{m}:= \frac{1}{N} \sum_{i=1}^N \left( a^\dagger_i\right)^m a_i^m  , \qquad m=2,3,\cdots.
	\label{r-charge}
\end{align}
Hence, thermal equilibrium states of this model in the micro-canonical ensemble are specified by $E=\langle H \rangle$ and $r_m := \langle R_{m} \rangle$.

We show that quantities such as $\langle \overline{x^m p^n } \rangle$ defined by \eqref{x-average} in the thermal equilibrium states can be determined by the bootstrap method.
By regarding the conserved charges \eqref{r-charge}, we modify the constraints \eqref{constraint-micro-canonical} for the micro-canonical ensemble as,
\begin{align}
	{\mathcal M}
	\succeq 0, \quad E= \langle H \rangle, 
	\quad  	\langle [ H ,O] \rangle=0,
	\quad r_m = \langle R_{m} \rangle, 
	\quad \langle O_1 O_2 \rangle	=\langle O_1 \rangle \langle O_2 \rangle.
	\label{constraint-GGE-HO}
\end{align}
What we should do is finding the possible values of $\langle \overline{x^m p^n } \rangle$  which are consistent with these constraints.
Actually, we can solve this problem analytically.
We define the operators $R_{mn}:=\frac{1}{N} \sum_{i=1}^N  \left( a^\dagger_i \right)^m \left( a_i \right)^n $ and substitute it to the constraint $\langle [ H ,O] \rangle=0$ in \eqref{constraint-GGE-HO}, and obtain
\begin{align}
	0= \langle [ H ,R_{mn}] \rangle = (m-n) \langle R_{mn} \rangle.
\end{align}
Hence, $\langle R_{mn} \rangle=0$ if $m \neq n$.
Since the quantities $\langle \overline{x^m p^n } \rangle$ can be expressed by a sum of $\langle R_{kl} \rangle$, the value of  $\langle \overline{x^m p^n } \rangle$  is determined by $\langle R_{11} \rangle=E/N-\hbar/2$ and $\langle R_{kk} \rangle=\langle R_{k} \rangle=r_k$.
For example, it is trivial but $\langle \overline{x^2 } \rangle$ becomes
\begin{align} 
	\langle \overline{x^2 } \rangle
	= \frac{1}{N} \sum_{i=1}^N 	\langle x_i^2  \rangle 
	= \frac{1}{N} \sum_{i=1}^N \hbar \left(  \langle  a^\dagger_i  a_i  \rangle + \frac{1}{2} \right)
	= \frac{E}{N}.
\end{align}
Similarly, the quantities $\langle \overline{x^m p^n } \rangle$  in this system in thermal equilibrium are completely fixed by the conserved charges $E$ and $\{ r_m \}$.

Note that we have used the constraint $\langle [ H ,O] \rangle=0$ and $r_m = \langle R_{m} \rangle$ only.
Here we comment on the other constraints in \eqref{constraint-GGE-HO}.
It is easy to show that the constraint $\langle O_1 O_2 \rangle	=\langle O_1 \rangle \langle O_2 \rangle$ is automatically satisfied in free particle systems, if we consider the averaged operators such as \eqref{x-average} and take $N$ large. 
(It implies that any stationary states of one-dimensional free particles are always thermal equilibrium states.)
Besides, the constraint ${\mathcal M} \succeq 0$ is satisfied, if we take appropriate conserved charge $E$ and $\{r_m \}$. Thus, this can be regarded as an ``initial value problem".\footnote{The issue of the particle species also reduces to the initial value problem. Namely, we need to prepare suitable charges, which are consistent with the given particle species.}
One way to generate suitable conserved charges is using a bose or fermi distribution function, which we employ in the next section.

\subsection{Example 2: Non-interacting $N$-anharmonic oscillator}
\label{sec-GGE-AHO}

Since the harmonic oscillator \eqref{H-HO-N} is so simple that we can solve it analytically, as a more nontrivial example, we consider non-interacting $N$-anharmonic oscillators,
\begin{align}
	H  =& \sum_{i=1}^N h_i, \quad h_i:= \frac{1}{2}  p_i^2  + \frac{1}{2} x_i^2 
	   + \frac{1}{4}  x_i^4 .
	\label{H-aho-N} 
\end{align}
Then, the system has the conserved charges,
\begin{align} 
	H^{(m)}:= \frac{1}{N} \sum_{i=1}^N h_i^m  , \quad m=1,2,3,\cdots.
	\label{GGE-charge}
\end{align}
Here $H^{(1)}$ is equivalent to the Hamiltonian \eqref{H-aho-N}.

By using the numerical bootstrap method, we seek the possible range of the expectation value of the operator
\begin{align}
	\overline{x^2} :=  \frac{1}{N} \sum_{i=1}^N  x^2_i
\label{x-GGE}
\end{align}
in this system at thermal equilibrium, as an example.
Hence, we derive the upper and lower bounds on $\langle \overline{x^2} \rangle$ under the constraints,
\begin{align}
	{\mathcal M}
	\succeq 0,
	\quad  	\langle [ H ,O] \rangle=0,
	\quad E^{(m)}:= \langle H^{(m)} \rangle, 
	\quad \langle O_1 O_2 \rangle	=\langle O_1 \rangle \langle O_2 \rangle.
	\label{constraint-GGE-AHO}
\end{align}
Here $\langle O_1 O_2 \rangle =\langle O_1 \rangle \langle O_2 \rangle$ is automatically satisfied at large-$N$ as we mentioned in the previous section.

We construct the bootstrap matrix ${\mathcal M}$ as follows.
Since we are interested in the averaged operator $ \overline{x^2} $ \eqref{x-GGE}, it is useful to take the seed operator
\begin{align}
	\tilde{O}_i:= \sum_{m=0}^{K_x} \sum_{n=0}^{K_p} b_{mn} x_i^m  p_i^n .
	\label{Otilde-GGE}
\end{align}
Then $\sum_{i=1}^N \langle \tilde{O}^{\dagger}_i \tilde{O}_i \rangle \ge 0  $ is satisfied for any $\{b_{mn} \}$, and we obtain the bootstrap matrix ${\mathcal M}$ \eqref{bootstrap-XP} where $\langle x^mp^n \rangle$ are replaced by $\langle \overline{x^mp^n} \rangle$.
In addition, $H^{(m)}$ can be expressed by $\overline{x^k p^l}  $ too.
Then, all the variables in the constraints \eqref{constraint-GGE-AHO} are expressed by the averaged operators $\overline{x^k p^l}  $, and we do not need to handle the operators for the individual particle such as $x_i$ and $p_j$ anymore.
In this way, the constraints \eqref{constraint-GGE-AHO} become formally equivalent to the constraints \eqref{constraint-statonary-E} for the single anharmonic oscillator with the additional constraints $ E^{(m)} = \langle H^{(m)} \rangle$ ($m=2,3,4,\cdots$) by identifying $\overline{x^mp^n}$ and $x^mp^n$.
This is a strong simplification, and is one advantage of the bootstrap analysis. 
(Related simplifications in the bootstrap method in multi-particle systems are expected, and several works on large-$N$ gauge theories have been done \cite{Han:2020bkb, Anderson:2016rcw, Lin:2020mme, Kazakov:2021lel, Kazakov:2022xuh}.)

Before solving the bootstrap problem \eqref{constraint-GGE-AHO}, we need to prepare suitable conserved charges $E$ and $ \{ E^{(m)} \}$.
For this purpose, we assume that the $N$ particles are all the same boson and they obey the standard bose distribution function with temperature $T=1/\beta$ and chemical potential $\mu$.\footnote{
	This system has infinite number of chemical potentials corresponding to the conserved charges \eqref{GGE-charge}. 
	If we take these chemical potentials zero except those corresponding to $E$ and $N$, we obtain the standard bose distribution function in \eqref{free-thermal-N}.
	Such a situation may be realized, if the particles interact each other very weakly.
}
Then the expectation value of the operator $\overline{x^2}$ in the thermal equilibrium state is given by
\begin{align} 
	\langle \overline{x^2} \rangle_{\beta, \mu}:= & 
	\frac{1}{N(\beta,\mu)}
	\sum_{n=0}^{\infty}
	\frac{1}{	 e^{\beta (e_n -\mu)}-1}
	\langle n | x^2 | n \rangle,
	\label{free-thermal-O}
	\\
	N(\beta,\mu):=                   & \sum_{n=0}^{\infty}
	\frac{1}{	 e^{\beta (e_n -\mu)}-1} .
	\label{free-thermal-N}
\end{align}
Here $| n \rangle$ is the energy eigenstate for the single particle and $e_n$ is its energy eigenvalue. $N(\beta,\mu)$ is the number of the bose particles. 
Then, through \eqref{GGE-charge}, we obtain,
\begin{align} 
	E^{(m)}(\beta,\mu) &= 	  
	\frac{1}{N(\beta,\mu)}
	\sum_{n=0}^{\infty}
	\frac{(e_n)^m}{	 e^{\beta (e_n -\mu)}-1}, \quad(m=1,2,\cdots).
	\label{higher-E}
\end{align}
In order to obtain the observables in the micro-canonical ensemble at given $E$ and $N$, we tune $T$ and $\mu$ such that $E=E(\beta,\mu)$ and $N=N(\beta,\mu)$ in the grand canonical ensemble.
Then, by using these tuned $T$ and $\mu$, we obtain $E^{(m)}(E,N)$ and $\langle \overline{x^2} \rangle(E,N) $.
Particularly, we will use $E^{(m)}(E,N)$ as the input of the bootstrap analysis in \eqref{constraint-GGE-AHO}, and test whether $\langle \overline{x^2} \rangle(E,N) $ is reproduced.

\begin{figure}
	\begin{tabular}{ccc}
		\begin{minipage}{0.33\hsize}
			\begin{center}
				\includegraphics[scale=0.6]{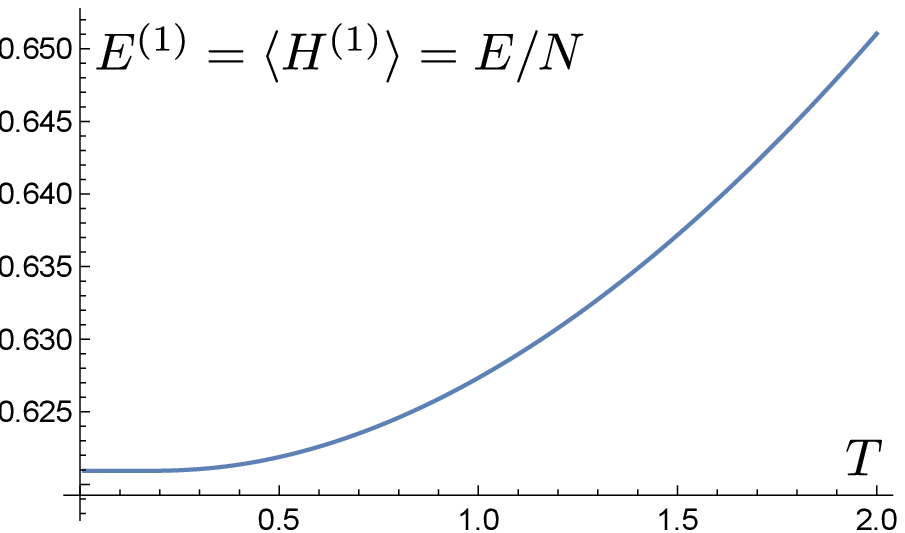}\\
				$T$ vs. $  E^{(1)}  $
			\end{center}
		\end{minipage}
		\begin{minipage}{0.33\hsize}
			\begin{center}
				\includegraphics[scale=0.6]{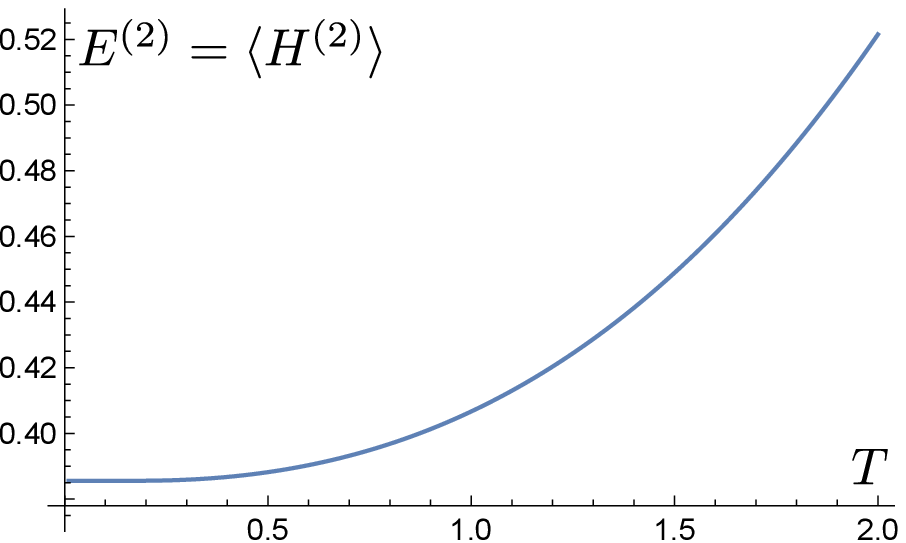}\\
				$T$ vs. $ E^{(2)} $
			\end{center}
		\end{minipage}
		\begin{minipage}{0.33\hsize}
			\begin{center}
				\includegraphics[scale=0.6]{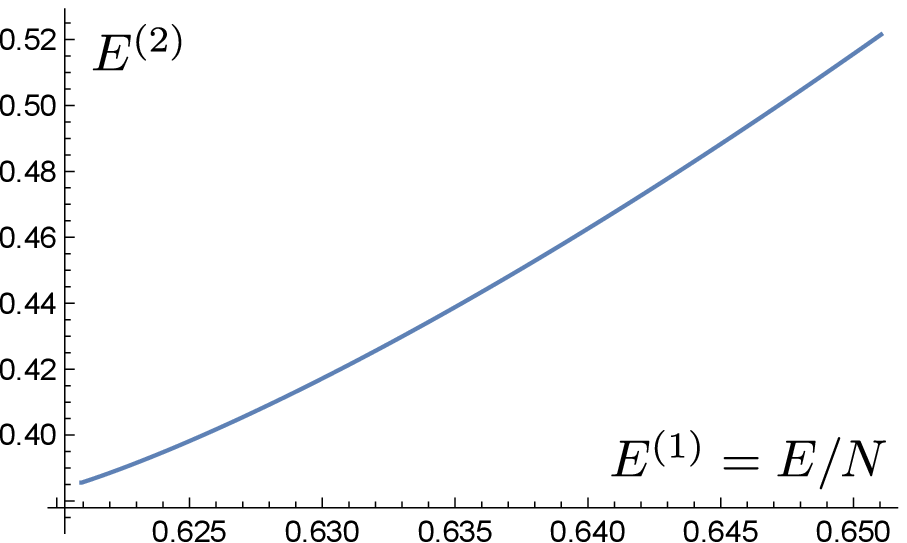}\\
				$ E^{(1)} $vs. $ E^{(2)} $
			\end{center}
		\end{minipage}
	\end{tabular}
	\caption{
		Temperature dependence of $ E^{(1)}=E/N   $ and $ E^{(2)} $ for the $N$ bose particles in the anharmonic oscillator \eqref{H-aho-N}. We take $N=100$.
		Through the plots of $T$ vs. $ E^{(1)}$ and  $T$ vs. $ E^{(2)}$, by eliminating $T$, we obtain the plot of $ E^{(1)}$ vs. $ E^{(2)}$.
	}
	\label{Fig-T-E}
\end{figure}

\begin{figure}
	\begin{center}
		\includegraphics[scale=0.7]{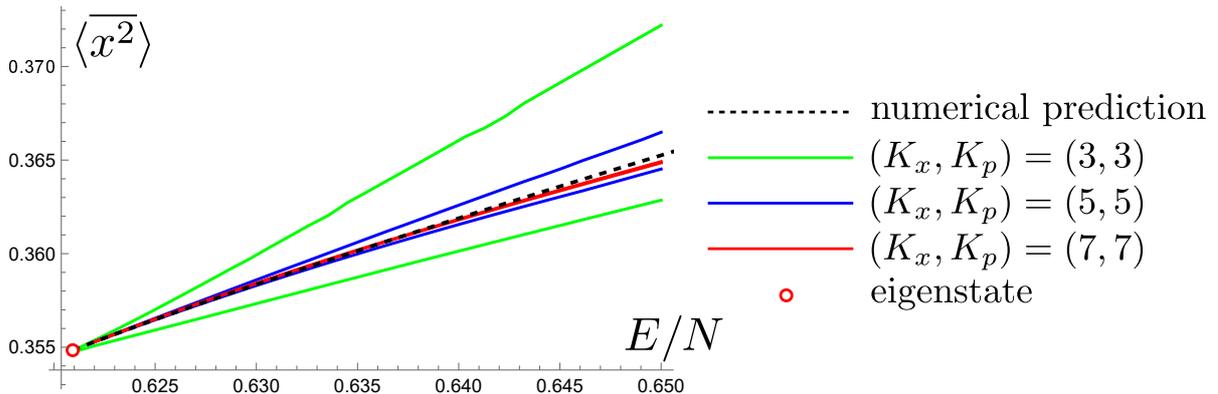}
	\end{center}
	\caption{
		$E$ vs. $ \langle \overline{x^2} \rangle $ for the $N$ bose particles in the anharmonic oscillator \eqref{H-aho-N}. We take $N=100$.
		For each $E$, we fix $E^{(2)}$ through the relation shown in Fig.~\ref{Fig-T-E} and use it as the input parameter in the bootstrap analysis \eqref{constraint-GGE-AHO}. 
		We observe that the bootstrap method reproduces the numerical prediction (the dashed line) as $(K_x,K_p)$ increase.
		Particularly, at $K_x=K_p=7$, the results of the bootstrap method are almost coincident with the numerical ones.
		However, the bootstrap method does not work properly in higher energy region $E/N > 0.65$.
	}
	\label{Fig-thermal-free}
\end{figure}

In our numerical analysis, we take $N=100$ and first fix $\mu(\beta,N)$ through \eqref{free-thermal-N} for each $\beta$.
Then, we compute the temperature dependence of $E$ and $E^{(2)}$ at $N=100$ as shown in Fig.~\ref{Fig-T-E}.
From these results, by eliminating temperature, we obtain $E^{(2)}(E,N)$ as shown in Fig.~\ref{Fig-T-E} (right panel).
Similarly, we plot $\langle\overline{ x^2 }\rangle(E,N)$ in Fig.~\ref{Fig-thermal-free}.

We perform the numerical bootstrap analysis by using this $E^{(m)}(E,N)$ in the constraint \eqref{constraint-GGE-AHO}.
Actually, we find that just $E^{(2)}$ is sufficient to reproduce $\langle\overline{ x^2 }\rangle(E,N)$ .
The results are shown in Fig.~\ref{Fig-thermal-free}.\footnote{
We solve this problem by using the Mathematica package ``SemidefiniteOptimization" with the option ``Mosek".
There, we imposed the parity condition $\langle\overline{ x^m p^n }\rangle=0$, ($n+m:$ odd) in order to perform the numerical analysis efficiently.
}
We find that the they are consistent with the thermal equilibrium state \eqref{free-thermal-O}.

However, the bootstrap method does not work properly in higher energy region $E/N > 0.65$.
Since the ground energy is $E/N = 0.6209$, the region, in which the bootstrap method works, is very low energy.
We guess that it may be a technical issue, and improvements in numerical analysis may resolve it.
Note that we can do in principle similar analysis for $N$ fermions by using the fermi distribution function in \eqref{free-thermal-N}.
However, the fermi energy is high at large $N$, and we need to handle higher energy, which would be difficult in our numerical bootstrap analysis.
Indeed, as far as we tried, we could not obtain reliable results.

{\normalsize 
\bibliographystyle{unsrt}
\bibliography{bBFSS} }

\end{document}